%% file: hoard_ms.tex
\documentclass[12pt,preprint]{aastex}

\shortauthors{Hoard et al.}
\shorttitle{Novalikes in the Infrared}

\slugcomment{The Astrophysical Journal, accepted 20 Mar 2014}

\begin{document}

\title{Novalike Cataclysmic Variables in the Infrared}
 
\author{D.~W.\ Hoard\altaffilmark{1,2,3}, 
Knox S.\ Long\altaffilmark{4},
Steve B.\ Howell\altaffilmark{5},
Stefanie Wachter\altaffilmark{2},
Carolyn S.\ Brinkworth\altaffilmark{6,7},
Christian Knigge\altaffilmark{8},
J.~E.\ Drew\altaffilmark{9},
Paula Szkody\altaffilmark{10},
S.\ Kafka\altaffilmark{11},
Kunegunda Belle\altaffilmark{12},
David R.\ Ciardi\altaffilmark{7},
Cynthia S.\ Froning\altaffilmark{13},
Gerard T.\ van Belle\altaffilmark{14},
and
M.\ L.\ Pretorius\altaffilmark{15}
}

\altaffiltext{1}{Eureka Scientific, Inc., Oakland, CA, USA}
\altaffiltext{2}{Max Planck Institut f\"{u}r Astronomie, Heidelberg, Germany}
\altaffiltext{3}{Visiting Scientist, MPIA; email:\ {\tt hoard@mpia.de}}
\altaffiltext{4}{Space Telescope Science Institute, Baltimore, MD, USA}
\altaffiltext{5}{NASA Ames Research Center, Moffett Field, CA, USA}
\altaffiltext{6}{{\it Spitzer} Science Center, California Institute of Technology, Pasadena, CA, USA}
\altaffiltext{7}{NASA Exoplanet Science Institute, California Institute of Technology, Pasadena, CA, USA}
\altaffiltext{8}{Physics \& Astronomy, University of Southampton, Southampton, UK}
\altaffiltext{9}{Centre for Astrophysics Research, Science \& Technology Research Institute, University of Hertfordshire, Hatfield, UK}
\altaffiltext{10}{Department of Astronomy, University of Washington, Seattle, WA, USA}
\altaffiltext{11}{Carnegie Institution of Washington, Department of Terrestrial Magnetism, Washington, DC, USA}
\altaffiltext{12}{Los Alamos National Laboratory, Los Alamos, NM, USA}
\altaffiltext{13}{Center for Astrophysics and Space Astronomy, Department of Astrophysical and Planetary Sciences, University of Colorado, Boulder, CO, USA}
\altaffiltext{14}{Lowell Observatory, Flagstaff, AZ, USA}
\altaffiltext{15}{Department of Physics, University of Oxford, Oxford, UK}

\begin{abstract}
Novalike cataclysmic variables have persistently high mass transfer rates and prominent steady state accretion disks.  We present an analysis of infrared observations of twelve novalikes obtained from the Two Micron All Sky Survey, the {\it Spitzer Space Telescope}, and the {\it Wide-field Infrared Survey Explorer} All Sky Survey.  The presence of an infrared excess at $\lambda\gtrsim3$--$5$~$\mu$m over the expectation of a theoretical steady state accretion disk is ubiquitous in our sample.  The strength of the infrared excess is not correlated with orbital period, but shows a statistically significant correlation (but shallow trend) with system inclination that might be partially (but not completely) linked to the increasing view of the cooler outer accretion disk and disk rim at higher inclinations.  We discuss the possible origin of the infrared excess in terms of emission from bremsstrahlung or circumbinary dust, with either mechanism facilitated by the mass outflows (e.g., disk wind/corona, accretion stream overflow, and so on) present in novalikes.  Our comparison of the relative advantages and disadvantages of either mechanism for explaining the observations suggests that the situation is rather ambiguous, largely circumstantial, and in need of stricter observational constraints.  
\end{abstract}


\section{Introduction}

Cataclysmic variables (CVs) are interacting binary stars containing an accreting white dwarf (WD) primary and a mass-losing, 
late-type secondary star that fills its Roche lobe.
CVs are a common pathway for binary star evolution that includes classical novae and possibly leads to the standard candles -- Type Ia supernovae -- that play a crucial role in modern cosmology (e.g., see \citealt{btn13}).  
In most CVs, accretion of matter from the secondary star onto the (non-magnetic) WD is mediated by a disk that extends close to the surface of the WD. 
CVs arguably represent the most observationally accessible disk-accreting astrophysical systems. They also include the closest examples of an accretion flow around a compact object and, hence, provide a basic laboratory for accretion disk physics that is relevant in fields ranging from star and planet formation to the central engines of quasars and active galactic nuclei.  See \citet{warner03} for a thorough review of CV types and observational characteristics.

All CVs vary in mean brightness on a number of characteristic timescales and amplitudes, but the overall nature of the variability probably reflects the time-averaged mass transfer rate from the secondary star.  Systems with mass transfer rates below a critical value ($\lesssim10^{-9}$~M$_{\odot}$~yr$^{-1}$ for orbital periods $\lesssim8$~hr; e.g., \citealt{patterson84}) show quasiperiodic outbursts of 3--5 mag (in visible light) that last days to weeks and recur on timescales of weeks to months to (in extreme cases) years.  These CVs are known as dwarf novae (DNe).
The outbursts of DNe are due to a thermal instability that converts the geometrically thin disk from a low temperature, mostly unionized state to a high temperature, ionized, optically thick state \citep{cannizzo88,osaki96,lasota01}. During a DN outburst, the mass transfer rate in the inner disk rises 
to $\gtrsim10^{-9}$~M$_{\odot}$~yr$^{-1}$ \citep{hameury98}. 

In contrast, CVs with persistently high mass transfer rates ($\gtrsim10^{-9}$~M$_{\odot}$~yr$^{-1}$; \citealt{patterson84,ballouz09}) remain in the high temperature state most of the time and have prominent, optically thick accretion disks that do not display outbursts.  These systems are known as novalikes (NLs). Because they are almost always in the high temperature state,
NLs provide a valuable opportunity to study a prototypical steady state accretion disk.

Disk-dominated CVs have been studied extensively in X-rays, the ultraviolet (UV), and visible wavelengths, where they exhibit different behavior depending on the sub-type and brightness state of the system.  A variety of components contribute to the spectral energy distributions (SEDs) of CVs, including the WD onto which matter is being accreted, the accretion disk itself, the boundary layer between the WD and the disk (mainly in the UV and shorter wavelengths), the interaction region where matter from the secondary star is entrained into the disk, and the secondary star itself. When the accretion rate is high, emission from the far-UV to the infrared (IR) is dominated by the accretion disk (e.g., see \citealt{hoard09} and Section~\ref{s:models}).

\citet{hoard09} used multi-wavelength archival and {\it Spitzer Space Telescope} Cycle-2 IR observations of the low-inclination (near face-on) NL V592~Cassiopeiae to construct an SED and system model from the UV to the IR.  They showed that there is an excess of flux density in the IR that is not reproduced by the usual complement of CV components (WD, accretion disk, secondary star).  They modeled this IR excess with a circumbinary dust disk.  

Through the gravitational torques that would be exerted on the central binary, circumbinary dust disks were proposed as an additional route for angular momentum loss driving the secular evolution of CVs \citep{spruit01,taam01,dubus02,taam03,belle04,willems05,willems07}.  However, \citet{hoard09} found that the implied mass of dust in V592~Cas ($\sim10^{21}$~g) was many orders of magnitude too small to be effective in that regard\footnote{Recent discoveries of cyclic eclipse timing variations interpreted as evidence of circumbinary planets around several CVs and pre-CVs (e.g., 
RR~Caeli -- \citealt{qian12a}; 
UZ~Fornacis -- \citealt{potter11}; 
DP~Leonis -- \citealt{qian10a,beuermann11}; 
NN~Serpentis -- \citealt{bmd06,pmb14,mpb14}; 
HW~Virginis -- \citealt{lee09}; 
NY~Virginis -- \citealt{qian12b}; 
QS~Virginis -- \citealt{parsons10,qian10b}) 
have possibly resurrected this scenario, since the added mass of a planet could increase the gravitational torques on the inner binary to levels sufficient to affect secular evolution. However, the veracity of the planet detections is still being debated in the literature 
(e.g., HU~Aquarii; see \citealt{schwarz09,qian11,horner11,godz12,hinse12,wittenmyer12}).  
Certainly the existence of circumbinary planets as a general class of object has now been firmly established by discoveries from {\it Kepler} (e.g., Kepler-34 and Kepler-35, \citealt{welsh12}; Kepler-16, \citealt{doyle11}; Kepler-38, \citealt{orosz12a}; Kepler-47, \citealt{orosz12b}), so the possibility that a CV could host a circumbinary planet should not be casually discarded.}. 
The importance of the contribution of a circumbinary dust disk to the SEDs of CVs (or even the presence of dust in CVs at all) is not yet firmly established nor universally accepted.
Alternative explanations for an IR excess in CVs have been proposed (e.g., bremsstrahlung; see \citealt{harrison13} and the discussion herein).

We report here on our {\it Spitzer} observations of a sample of eleven additional NLs.
These observations were obtained in order to study the general properties of CVs in the mid-IR; NLs were selected as targets to ensure the presence of bright accretion disks and avoid the complications imposed by DN outbursts or strong magnetic fields in other types of CV.
At the same time, 
these NLs offer an opportunity to explore 
the potential observational consequences of 
the mass outflows from the inner binary 
expected in high mass transfer rate systems 
(e.g., disk corona/wind, stream overflow, and so on).
Because these mechanisms for relocating matter out of the WD Roche lobe and into circumbinary space operate strongly in NLs, these CVs are prime candidates for probing for the presence of dust.

\section{Targets and Observations}

Our primary targets are the nine brightest (in visible/near-IR light) NLs, which were observed in our {\it Spitzer} Cycle-5 program (50068). From brightest to faintest in Ks-band, they are:\ IX~Velorum, V3885~Sagittarii, RW~Sextantis, QU~Carinae, TT~Arietis, RW~Trianguli, UX~Ursae~Majoris, V347~Puppis, and VY~Sculptoris.
In addition, we utilized a serendipitous observation of UX~UMa from {\it Spitzer} program 40204.
The bulk of our data for IX~Vel was actually obtained as part of our Cycle-2 program (20221), with only a repeat spectroscopic measurement during Cycle-5.
We also include two additional NLs from our Cycle-2 program, WX~Arietis and V442~Ophiuchi.  Although fainter than the Cycle-5 program targets, these objects are intermediate inclination members of the SW~Sextantis sub-type of CV (as is TT~Ari, and probably also RW~Tri, UX~UMa, and V347~Pup).  The SW~Sex stars are NLs that share a number of unusual observational characteristics \citep{honeycutt86,szkody90,thor91} that are likely linked to the presence of a self-occulting, flared accretion disk with a bright spot at the impact site of the matter stream from the secondary star with the disk edge \citep{hoard03,hlk10,dsm13}\footnote{Also see D.~W.\ Hoard's Big List of SW~Sextantis Stars at \url{http://www.dwhoard.com/biglist}.}. 
Although not re-observed as part of our {\it Spitzer} Cycle-5 program, V592~Cas from our Cycle-2 program is included here as a baseline for comparison to a CV with known IR excess.  

These targets span a wide range of orbital period and inclination (see Table~\ref{t:targets}). 
To first order, longer orbital period correlates with higher mass transfer rate from the secondary star \citep{patterson84,hnr01}, in principle allowing us to probe the effect of differing rates of mass flow through the accretion disk and self-irradiation via accretion-generated luminosity close to the WD. The lower inclination (near face-on) CVs provide a view of the entire radial profile of the accretion disk, as well as the maximum projected disk surface area.  Thus, they potentially offer the most sensitive probes of departures of the disk SED from the steady state prediction.

Table~\ref{t:log} lists the log of {\it Spitzer} observations and other archival data for each target.  We utilized all of {\it Spitzer}'s instruments during this observing campaign.  Depending on target brightness, this provided photometric data in up to 6 bands from the Infrared Array Camera (IRAC; IRAC-1--IRAC-4 bands at 3.55, 4.5, 5.7, 7.9~$\mu$m), the Infrared Spectrograph Peak-Up Imaging array (IRSPUI; blue or red band at 15.8 or 22.3~$\mu$m), and the Multi-band Imaging Photometer for {\it Spitzer} (MIPS; MIPS-24 band at 23.7~$\mu$m)\footnote{See the corresponding Instrument Handbooks for details of each instrument, \url{http://irsa.ipac.caltech.edu/data/SPITZER/docs/irac/iracinstrumenthandbook/}, \url{http://irsa.ipac.caltech.edu/data/SPITZER/docs/irs/irsinstrumenthandbook/}, and \url{http://irsa.ipac.caltech.edu/data/SPITZER/docs/mips/mipsinstrumenthandbook}.}.  For seven of the targets, we also obtained Infrared Spectrograph low resolution spectra, spanning wavelengths from $\approx5$~$\mu$m to (depending on target brightness) $\approx8$, 14, or 20~$\mu$m.  

In addition to the {\it Spitzer} data, we utilized IR photometry from the Two-Micron All Sky Survey (2MASS, JHKs bands; \citealt{skrutskie06}) and the {\it Wide-field Infrared Survey Explorer} ({\it WISE}) All Sky Survey \citep{wright10}.  {\it WISE} provides photometric data in four bands, denoted W1, W2, W3, and W4, at wavelengths of 
3.35, 4.6, 11.6, and 22.1~$\mu$m, respectively.
When available, the {\it Spitzer} observations generally reach greater sensitivity (hence, improved photometric accuracy) compared to {\it WISE\/}\footnote{The nominal $5\sigma$ point source sensitivity limits for the {\it WISE} All Sky Survey are 0.08, 0.11, 1, and 6 mJy in W1, W2, W3, and W4, respectively \citep{wright10}; cold {\it Spitzer} surpassed these sensitivity limits for IRAC-1 ($\approx$W1), IRAC-2 ($\approx$W2), and MIPS-24 ($\approx$W4) in total on-source times of less than a minute (see the {\it Spitzer} Performance Estimation Tool at \url{http://ssc.spitzer.caltech.edu/warmmission/propkit/pet/senspet/index.html}).}.  The latter are useful to provide confirmation of the {\it Spitzer} photometry, a second epoch measurement, and/or coverage at wavelengths not sampled by {\it Spitzer} observations.

\subsection{Photometric and Spectroscopic Data Extraction}
\label{s:extractions}

\begin{figure*}[tb]
\epsscale{0.75}
\plotone{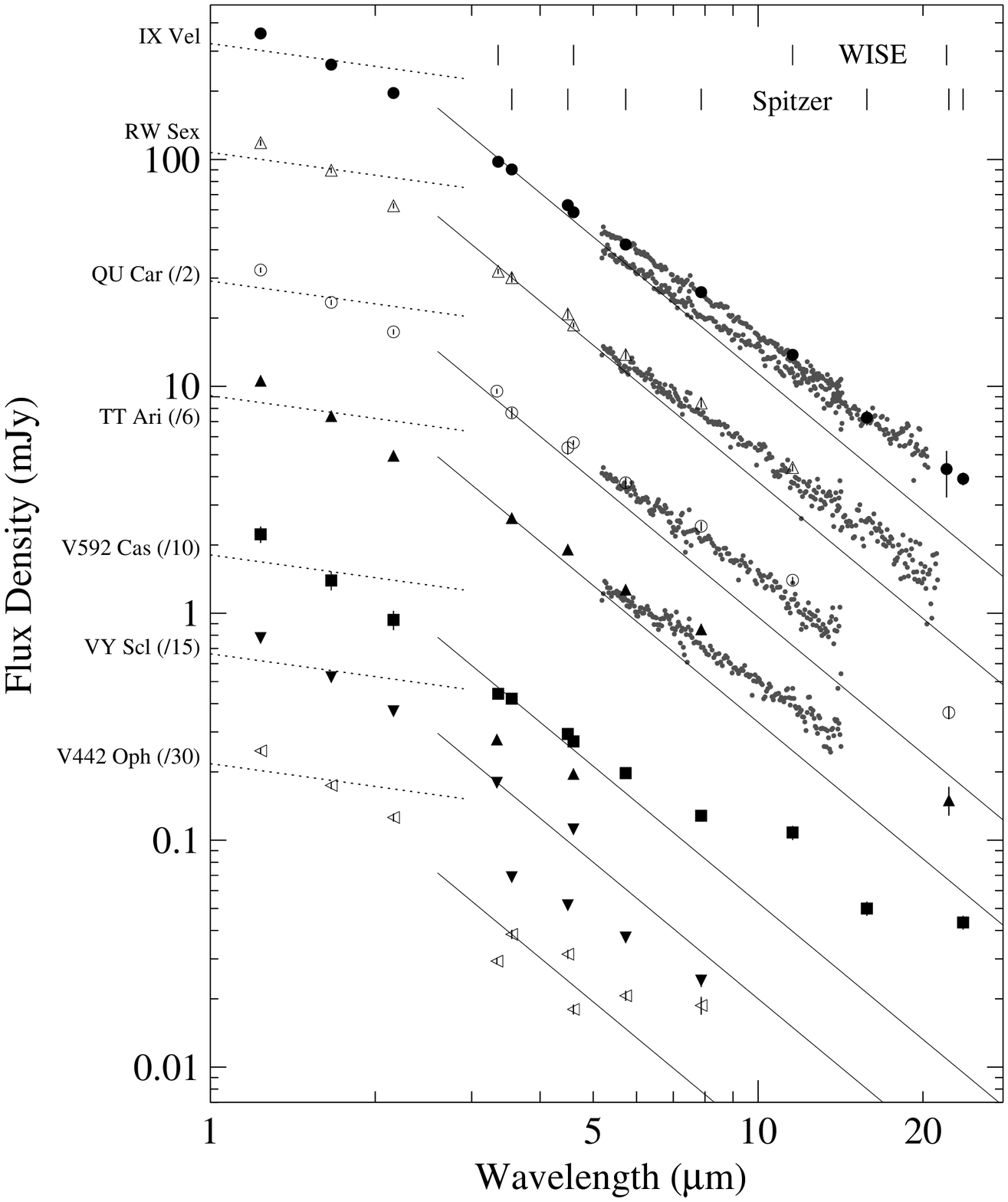}
\epsscale{1.0}
\caption{
IR SEDs of NLs, constructed from 2MASS, {\it WISE}, and {\it Spitzer} data.  Isophotal wavelengths of the {\it WISE} and {\it Spitzer} photometric bands are indicated.  The {\it Spitzer} spectroscopic data are shown as small, gray filled circles.  The photometric data are symbol-coded for each CV and, in some cases, have been offset via a constant scale factor for clarity.  
From top to bottom, the data correspond to 
IX~Vel (filled circles),
RW~Sex (unfilled upward triangles),
QU~Car (unfilled circles; scale factor of 1/2),
TT~Ari (filled upward triangles; scale factor of 1/6),
V592~Cas (filled squares; scale factor of 1/10),
VY~Scl (filled downward triangles; scale factor of 1/15),
and
V442~Oph (unfilled left-pointing triangles; scale factor of 1/30).
$1\sigma$ error bars are plotted on all of the photometric data, but in some cases are smaller than the plot symbols.
The dotted lines show the ``disk spectrum'' portion of the steady state accretion disk SED ($f_{\nu}\propto\lambda^{-1/3}$; \citealt{FKR02}), normalized to the mean flux density of the three 2MASS bands.
The solid lines show the Rayleigh-Jeans tail portion of the accretion disk SED ($f_{\nu}\propto\lambda^{-2}$; \citealt{FKR02}), normalized to the {\it Spitzer} IRAC-1 (3.55 $\mu$m) flux density.
\label{f:seds01uncorrected}}
\end{figure*}

\begin{figure*}[tb]
\epsscale{0.75}
\plotone{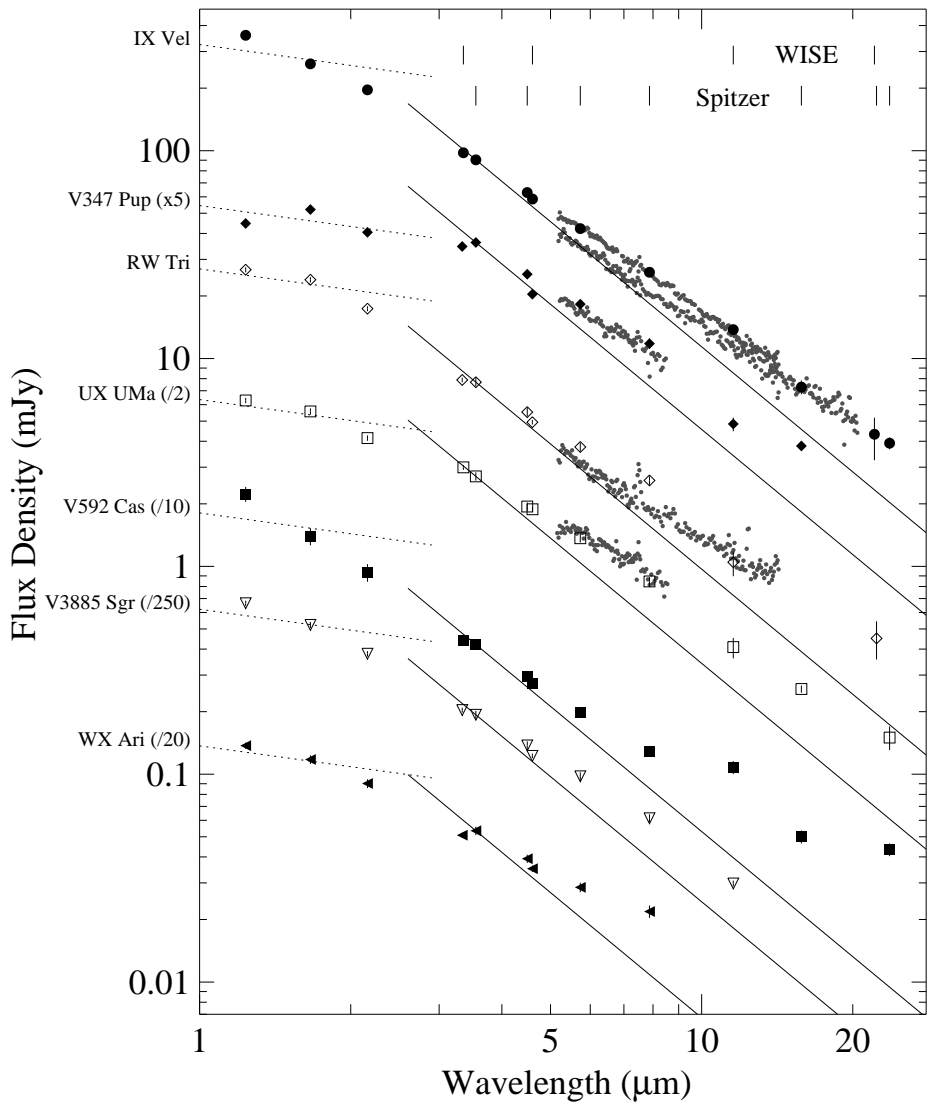}
\epsscale{1.0}
\caption{
As in Figure~\ref{f:seds01uncorrected}, for additional NLs.  
From top to bottom, the data correspond to 
IX~Vel (filled circles),
V347~Pup (filled diamonds; scale factor of 5),
RW~Tri (unfilled diamonds),
UX~UMa (unfilled squares; scale factor of 1/2),
V592~Cas (filled squares; scale factor of 1/10),
V3885~Sgr (unfilled downward triangles; scale factor of 1/250),
and
WX~Ari (filled left-pointing triangles; scale factor of 1/20).
IX~Vel and V592~Cas are repeated from Figure~\ref{f:seds01uncorrected} for comparison.
\label{f:seds02uncorrected}}
\end{figure*}

In most cases, we utilized aperture photometry performed with IRAF\footnote{IRAF is distributed by the National Optical Astronomy Observatory, which is operated by the Association of Universities for Research in Astronomy (AURA) under cooperative agreement with the National Science Foundation.} (version 2.16; method as described in \citealt{hoard09}) to extract the {\it Spitzer} photometry from the corrected basic calibrated data (CBCD) IRAC images, the standard BCD IRSPUI images, and the enhanced BCD (EBCD) MIPS images.  We generally utilized the smallest calibrated aperture case whenever possible (as described in the corresponding Instrument Handbooks).  We also performed aperture and PRF-fitting photometry with MOPEX/APEX\footnote{See \url{http://irsa.ipac.caltech.edu/data/SPITZER/docs/dataanalysistools/tools/mopex/mopexusersguide/}.} as a check of the IRAF results for IRAC and IRSPUI whenever the presence of nearby sources might have affected the aperture photometry.  We generally preferred MOPEX/APEX aperture photometry for the MIPS observations in order to fully utilize the mosaicking properties of the EBCD images. 
The final photometry, along with the method used to procure it from each instrument for each target, are listed in Table~\ref{t:phot} 
(and plotted in Figures~\ref{f:seds01uncorrected} and \ref{f:seds02uncorrected}; see discussion in Section~\ref{s:ir_excess}); 
pertinent notes are described on a target-by-target basis below.
We transformed the cataloged 2MASS and {\it WISE} photometry from magnitudes to flux densities using the zeropoint information in \citet{cohen03} and 
\citet{jcm11}, respectively.

For the targets with {\it Spitzer} IRS observations, we extracted spectra from the post-BCD pipeline images using the standard procedure with the SPICE\footnote{See \url{http://irsa.ipac.caltech.edu/data/SPITZER/docs/dataanalysistools/tools/spice/spiceusersguide/}.} software (version 2.3F).  Spectra were extracted from each nod individually, then combined using a weighted average.  For plotting purposes, the spectroscopic data were block-averaged in wavelength bins of two points.

Because CVs are intrinsically variable objects, one must 
be cautious when combining multiple data sets obtained at (significantly) different times.  
The NLs studied here, in particular, are subject to stochastic brightness variations with amplitude up to several tenths of a magnitude across a wide range of time scales, as well as transitions to low mass transfer states resulting in sustained drops in overall brightness (see Section~\ref{s:app_vyscl}).
Whenever possible, we attempted to mitigate the effects of variability between data sets by consulting available brightness information from published sources or the American Association of Variable Star Observers (AAVSO) Light Curve Generator\footnote{See \url{http://www.aavso.org/lcg}.}, or by comparing overlapping data obtained at similar wavelengths but different times.  
Unfortunately, there is no {\it ex post facto} means of precisely evaluating the relative effect of stochastic brightness variations on approximately orbital time scales in non-simultaneous archival data sets.  However, the amplitude of this type of variability is substantially smaller than that of the long-term brightness state changes in NLs and, to some extent, it should ``average out'' when utilizing multiple data points sampled over intervals up to a few orbits in duration.

When brightness offsets between overlapping data sets existed, we determined a correction to the fainter data to bring it into agreement with the brighter data.
This was accomplished by calculating a Rayleigh-Jeans-like (RJL) offset function, in which an additive offset is used to match the shortest wavelength point in the faint data to the corresponding point in the bright data, then scaling this offset by $\lambda^{-2}$ at each longer wavelength point (i.e., the offset function is $\Delta f_{\nu}(\lambda) = \Delta f_{\nu,0} \, \lambda^{-2}$).  
Physically, this correction implicitly assumes that the variation in total system brightness is primarily caused by a change in the accretion disk, and that the disk dominates the system brightness in the IR -- 
we discuss this in more detail in Section~\ref{s:ixvel_model}.
We have applied offsets conservatively, restricting this procedure only to cases for which there is a clear and significant discrepancy between overlapping data sets and/or an independent confirmation of a change in mean system brightness at one of our observation epochs.
Specific details of corrections applied to data sets on a target-by-target basis are provided below.
The corrected data are plotted in Figures~\ref{f:seds01} and \ref{f:seds02} (see discussion in Section~\ref{s:ir_excess}).

\subsection{Data Processing Notes for Individual Targets}
\label{s:notes}

The following sections describe any specific issues that arose during the data processing for each target.  In addition, we examined the {\it WISE} All Sky Atlas images for each target; except as noted below for V592~Cas and RW~Tri, we did not find any anomalous features (either real or due to artifacts) that would affect the cataloged {\it WISE} photometry.

\subsubsection{V592 Cas}
\label{s:v592cas_data}

For this work, we utilized the final data processing from the {\it Spitzer} Heritage Archive\footnote{See \url{http://sha.ipac.caltech.edu}.} to re-extract the IR photometry for V592~Cas, as well as utilizing the {\it WISE} All Sky Survey photometry, which were not available for \citet{hoard09}.  Flux densities are listed in Table~\ref{t:phot}; for the most part, the new {\it Spitzer} values differ only slightly ($\lesssim1$\%) from those given in \citet{hoard09}.  The exceptions are the IRAC-4 (7.9~$\mu$m) and IRSPUI (15.8~$\mu$m) points, which differ by $+5$\% and $-10$\%, respectively.  This is consistent with revisions to the final flux calibration and aperture correction values made for the {\it Spitzer} end-of-cryo-mission data archive \citep{cih12}.

\citet{harrison13} independently re-derived the {\it Spitzer} photometry for V592~Cas as part of their examination of the {\it Spitzer} photometry and {\it Herschel Space Observatory} upper limits for several CVs.  They focused on magnetic CVs and DNe, and only V592~Cas is in common between their target sample and ours.  They obtained essentially the same photometric values as ours, and also noted the $\approx10$\% decrease in the revised IRSPUI flux density measurement.

The cataloged {\it WISE} All Sky W3 (11.6~$\mu$m) flux density for V592~Cas ($1.08\pm0.08$~mJy) is anomalously bright compared to the overall trend in its SED (see Figures~\ref{f:seds01uncorrected} and \ref{f:seds02uncorrected}).
If real, this would be an important finding, since it could indicate the presence of the 10-$\mu$m silicate emission feature, which is a hallmark of circumstellar dust around isolated WD stars \citep{jura07,jura09,reach09} and is also observed due to dust in the ejecta of post-outburst classical and recurrent novae \citep{evans07}.
Such a feature is not seen in the IRS spectra of our other targets; unfortunately, V592~Cas has no IRS spectral observation with which to verify its absence.

The W3 photometry could be contaminated by the brighter neighbor (designated WISE J002053.73+554219.3) that is located 13\arcsec\/ east of V592~Cas.
\citet{harrison13} suggest that the large full-width-at-half-maximum (FWHM) of the W3 point-spread-function (PSF) ensured that the W3 photometry of V592~Cas was contaminated.  Although this explanation is ultimately correct in its conclusion (see below), it is not entirely accurate:\ 
the {\it WISE} W1 and W2 photometry of V592~Cas agrees closely with the adjacent IRAC-1 and IRAC-2 points, so do not appear to be contaminated by the neighboring source, despite the fact that the W1 and W2 PSF FWHMs ($6.1\arcsec\times5.6\arcsec$ and $6.8\arcsec\times6.1\arcsec$, respectively) are comparable in size to that of W3 ($7.4\arcsec\times6.1\arcsec$)\footnote{See \url{http://wise2.ipac.caltech.edu/docs/release/allsky/expsup/sec4\_4c.html\#coadd\_psf}.}.
The crucial factor is that the wings of the W3 PSF are brighter relative to the peak at larger distances from the center compared to the W1 and W2 PSFs; 
at 13\arcsec\/ from the center, the W3 PSF wings are $\sim5$\% of the peak vs.\ $\sim1$\% of the peak at the same location in the W1 and W2 PSFs.

Given the potential importance of an elevated W3 flux density in V592~Cas, we thoroughly examined the 
possibility of
contamination from the neighboring source.  We downloaded the {\it WISE} All Sky Release Atlas images of the V592~Cas field from the Infrared Science Archive (IRSA)\footnote{See \url{http://irsa.ipac.caltech.edu/index.html}.}.  We used the bright isolated nearby star WISE~J002104.75+554356.9 (W2=8.22$\pm$0.02~mag, W3=8.12$\pm$0.02~mag) to define a local PSF, then used the IRAF implementation of the DAOPHOT package \citep{stetson87} to extract PSF-fit photometry for V592~Cas and its nearby neighbor from the W2 and W3 Atlas images.  Figure~\ref{f:v592cas_psfsub} shows the W3 Atlas image of V592~Cas, along with the PSF subtractions of the bright neighbor only and both stars together. 

The {\it WISE} All Sky Survey catalog photometry for the neighbor is W2=9.80(2)~mag and W3=9.59(3)~mag; for V592~Cas, the corresponding photometry is W2=12.00(2)~mag and W3=11.17(8)~mag.  We obtain values of W2=9.83(3)~mag and W3=9.64(3)~mag for the neighbor, in agreement with the {\it WISE} All Sky Survey values (within $1\sigma$).  For V592~Cas, we obtain W2=11.99(3)~mag and W3=11.58(3)~mag; although our W2 value is in agreement with the {\it WISE} All Sky Survey value, our W3 value is significantly fainter, corresponding to a flux density of 0.74(2)~mJy.  
This value is consistent with the predicted value based on a full SED model (see Section~\ref{s:v592cas_model}) and is used throughout the rest of this work (e.g., see Figures~\ref{f:seds01} and \ref{f:seds02}).  
So, unfortunately, it appears that there is no enhanced W3-band emission in V592~Cas.
The target with the next brightest W3 flux density is QU~Car (see Section~\ref{s:qucar_data}), but our IRS spectrum of this target rules out the presence of the 10~$\mu$m silicate emission feature.
See Section~\ref{s:missing} for additional discussion of the ``missing'' silicate emission feature in CVs.

\subsubsection{IX Vel}
\label{s:ixvel_data}


Our {\it Spitzer} Cycle-5 IRS spectrum of IX~Vel is fainter than the Cycle-2 IRS spectrum and IRAC photometry.  Unfortunately, we found no useful published or AAVSO brightness state data to correlate with the timing of the {\it Spitzer} observations.  
This would have enabled us to independently estimate a relative offset between the two IRS spectra.
Instead, to compensate for the difference in brightness between the two spectra, we added a RJL offset (as described in Section~\ref{s:extractions}) to the Cycle-5 spectrum, with initial value of +8.2 mJy at the short wavelength ($5.2$~$\mu$m) end of the Cycle-5 spectrum.
This brings the Cycle-5 spectrum into agreement with the Cycle-2 spectrum.
There is also a Cycle-5 IRS Short-High (high resolution) spectrum of IX~Vel, which is not used here.  We note that this spectrum matches the continuum shape of the corresponding IRS low resolution spectra, but is considerably noisier and does not show strong emission or absorption features.

\subsubsection{QU Car}
\label{s:qucar_data}


Observations from the AAVSO and \citet{kafka12}
show that QU~Car was variable between $V\approx11.2$--11.8~mag during our {\it Spitzer} observations. Although the IRS spectrum is slightly faint compared to the {\it Spitzer} and {\it WISE} photometry, there are no significant discrepancies between the various data sets.

\subsubsection{RW Sex}


The long-term AAVSO light curve for RW~Sex shows no significant changes in mean brightness level between the observation dates of the 2MASS, {\it Spitzer}, and {\it WISE} data sets.

\subsubsection{TT Ari}
\label{s:ttari_data}


The AAVSO light curve of TT~Ari shows no significant optical brightness change between the 2MASS and {\it Spitzer} observations; however, the {\it WISE} photometry was obtained during a faint state between JD 2455100--2455600.  During this time, TT~Ari was at $V\sim15.5$~mag instead of its normal $V\sim10.75$~mag. 
To compensate, we applied a RJL offset 
(as described in Section~\ref{s:extractions})
to the {\it WISE} photometry, 
with initial value of +16.5~mJy at W1 (3.35~$\mu$m).
Unlike the case of IX~Vel, for which we had a second, bright spectrum in an overlapping wavelength interval with which to estimate the correct offset function for the faint spectrum, we do not have photometry at exactly overlapping wavelengths to compare for TT~Ari.  Instead, we set the initial value of the offset function such that the corrected {\it WISE} photometry falls along the Rayleigh-Jeans tail accretion disk spectrum normalized to the IRAC-1 value (see Section~\ref{s:ir_excess}).
Although there is some concern that the SED might manifest a significant non-RJL secondary star contribution during such a deep faint state, the RJL offset does not appear to introduce a significantly discrepant slope to the corrected {\it WISE} photometry (see Figure~\ref{f:seds01}).

\subsubsection{RW Tri}


The long-term AAVSO light curve of RW Tri shows that there are no significant changes in the mean optical brightness level between the dates of the 2MASS, {\it Spitzer}, and {\it WISE} observations.
However, the IRS spectrum is fainter than the overlapping IRAC-3 and IRAC-4 points.
To compensate, we applied a RJL offset 
(as described in Section~\ref{s:extractions})
to the IRS and IRSPUI data (the latter were obtained contemporaneously with the former), with initial value of +0.7~mJy at $5.2$~$\mu$m.
The corrected IRS data are in better overall agreement with the other data, albeit still slightly fainter than IRAC-4 and slightly brighter than W3.
A faint, linear diffraction artifact passes near RW~Tri in the {\it WISE} W3 image; however, the surface brightness of the artifact is $\lesssim5$\% of the local background.  Consequently, it does not appear to have had any significant effect on the corresponding photometry.

\subsubsection{V347 Pup}


The AAVSO light curve coverage of V347~Pup stops in late 2002; prior data show variability of $\sim1$~mag (full amplitude) on timescales of hundreds of days, around an approximately constant mean brightness ($V\sim 13.5$~mag).  
The {\it WISE} photometry is systematically fainter than the {\it Spitzer} data.  
To compensate, we applied a RJL offset 
(as described in Section~\ref{s:extractions})
to the {\it WISE} photometry, with initial value of +1.2~mJy at W1 (3.35~$\mu$m).
The 2MASS photometry also seems faint compared to the {\it Spitzer} data, but we have no corroborating data with which to determine an appropriate offset.

\subsubsection{UX UMa}


The AAVSO data for UX~UMa show no significant changes in mean brightness level between the observation dates of the 2MASS, {\it Spitzer}, and {\it WISE} data sets.

\subsubsection{VY Scl}
\label{s:vyscl}


The 2MASS observations of VY~Scl were obtained during a normal bright state ($V\approx13.5$~mag).  Recent photometric coverage is sparse, but the {\it WISE} photometry were likely also obtained during a normal bright state, as they appear to be consistent with the extrapolation in wavelength of the 2MASS values.  However, the {\it Spitzer} Cycle-5 observations were obtained during the transition from a low state ($V\sim18$~mag) back to the bright state.  Using the optical light curves in \citet{greiner10} and from the AAVSO, we estimate $V\sim15.5$~mag during the IRS observations and $V\sim14.6$~mag during the IRAC observations.  
To compensate, we applied a RJL offset (as described in Section~\ref{s:extractions}) to the IRAC photometry, with initial value of +1.35~mJy at IRAC-1 (3.55~$\mu$m).  
The IRSPUI point, which was obtained when VY Scl was $\Delta V\sim0.9$~mag fainter than during the IRAC observations, was offset by +0.1 mJy, corresponding to an RJL offset with initial value of +1.9~mJy at IRAC-1.
The Cycle-5 IRS spectrum of VY~Scl is very noisy, and is not used or plotted here, although we note that it is roughly consistent with the other {\it Spitzer} photometry.

\subsubsection{V3885 Sgr}


The AAVSO long-term light curve of V3885~Sgr is stable through the end of coverage in late 2005.  The assembled SED shows no obvious signs of flux density level offsets between the various data sets.

\subsubsection{V442 Oph}

 
\citet{ballouz09} summarize the optical variability history of V442~Oph, which ranges from $V=12.6$~mag to $V=15.5$~mag, although the typical faint state level is $V=14.0$~mag.  Unfortunately, the AAVSO coverage of V442~Oph is very sparse, with only a handful of validated data points in the past 2 decades.  The {\it WISE} data are fainter than the corresponding IRAC photometry.  To compensate, we applied a RJL offset (as described in Section~\ref{s:extractions}) to the {\it WISE} photometry, with initial value of +0.4~mJy at W1 (3.35~$\mu$m).  The offset W2 flux density is still slightly fainter than IRAC-2.

\subsubsection{WX Ari}

 
There are no AAVSO data for WX~Ari during 1999--2009.  It was in a faint state ($V\sim17.5$~mag) below its normal brightness level ($V\sim15$~mag) from late-2012 until early-2013, but the decline to this state does not appear to have started until late-2011 and would not have affected any of the data used here.

\section{Analysis}

In this section we will 
establish that the IR SEDs of these NLs deviate from the expectation of the standard steady state accretion disk theory (Section~\ref{s:ir_excess}),
correlate the strength of this deviation with two fundamental system parameters (inclination and orbital period; Section~\ref{s:syspars}),
and
construct physically realistic, multi-component models to reproduce the observed SEDs of several of our targets (Section~\ref{s:models}).

\subsection{Infrared Excess}
\label{s:ir_excess}

\citet{FKR02} review the theoretical spectrum of a steady state CV accretion disk, and divide it into regions of the Wien exponential, which is dominated by the hottest inner regions of the disk ($T \sim T_{\rm wd}$), the ``characteristic disk spectrum'' at intermediate temperatures, which has $f_{\nu}\propto\lambda^{-1/3}$, and the Rayleigh-Jeans tail for the coolest disk regions, which is characterized by $f_{\nu}\propto\lambda^{-2}$.  NLs, in particular, are expected to have accretion disks that conform to this prescription, although it has been known for some time that the observed NL disk spectra tend to be redder than predicted by the standard model at short wavelengths corresponding to UV emission from the hot, inner disk (e.g., \citealt{wade88}).  This is often interpreted as evidence for a truncation of the inner disk -- so that the expected hottest regions are absent and do not contribute to the observed spectrum -- but the exact reason for the discrepancy is currently uncertain.  

\begin{figure*}[tb]
\epsscale{0.75}
\plotone{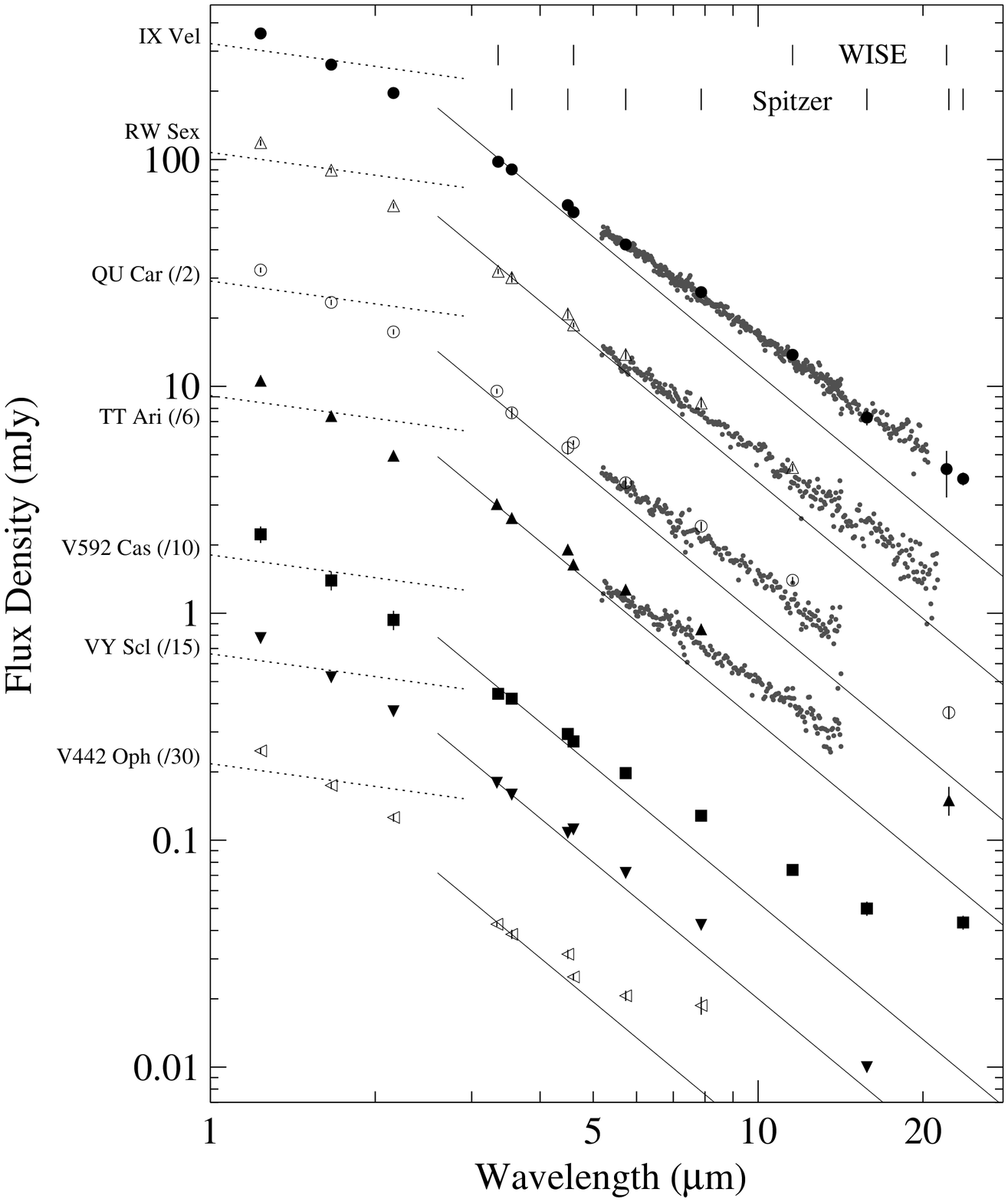}
\epsscale{1.0}
\caption{
As in Figure~\ref{f:seds01uncorrected}, but using the RJL-offset-corrected data where applicable (see Section~\ref{s:extractions} and individual target notes in Section~\ref{s:notes}) and the PSF-fit photometry value for the W3 point of V592~Cas (see Section~\ref{s:v592cas_data}).
\label{f:seds01}}
\end{figure*}

We are focussing here on the IR wavelengths corresponding to the cooler, outer accretion disk.
Before moving on to consideration of the mid-IR region of the SEDs, we first fit 
a characteristic disk spectrum to just the near-IR 2MASS data for each of our targets, 
normalized to the mean flux density of the three bands.  
For many of the targets, the JHKs data points are linearly decreasing (in log-log space), and have a steeper (negative) slope than the characteristic disk spectrum model, implying that the actual $f_{\nu}\propto\lambda^{-1/3}$ region occurs at shorter wavelengths, and the near-IR is already transitioning into the Rayleigh-Jeans tail spectrum.
\citet{szkody77} noticed a similar trend in pioneering near-IR (JHK) observations of six DNe obtained at or near maximum light, along with TT~Ari (whose nature at the time was uncertain but believed to be related to the DNe).
The targets displaying this behavior in our investigation are IX~Vel, RW~Sex, QU~Car, TT~Ari, V592~Cas, VY~Scl, and V442~Oph (i.e., the targets shown in Figures~\ref{f:seds01uncorrected} and \ref{f:seds01}).

\begin{figure*}[tb]
\epsscale{0.75}
\plotone{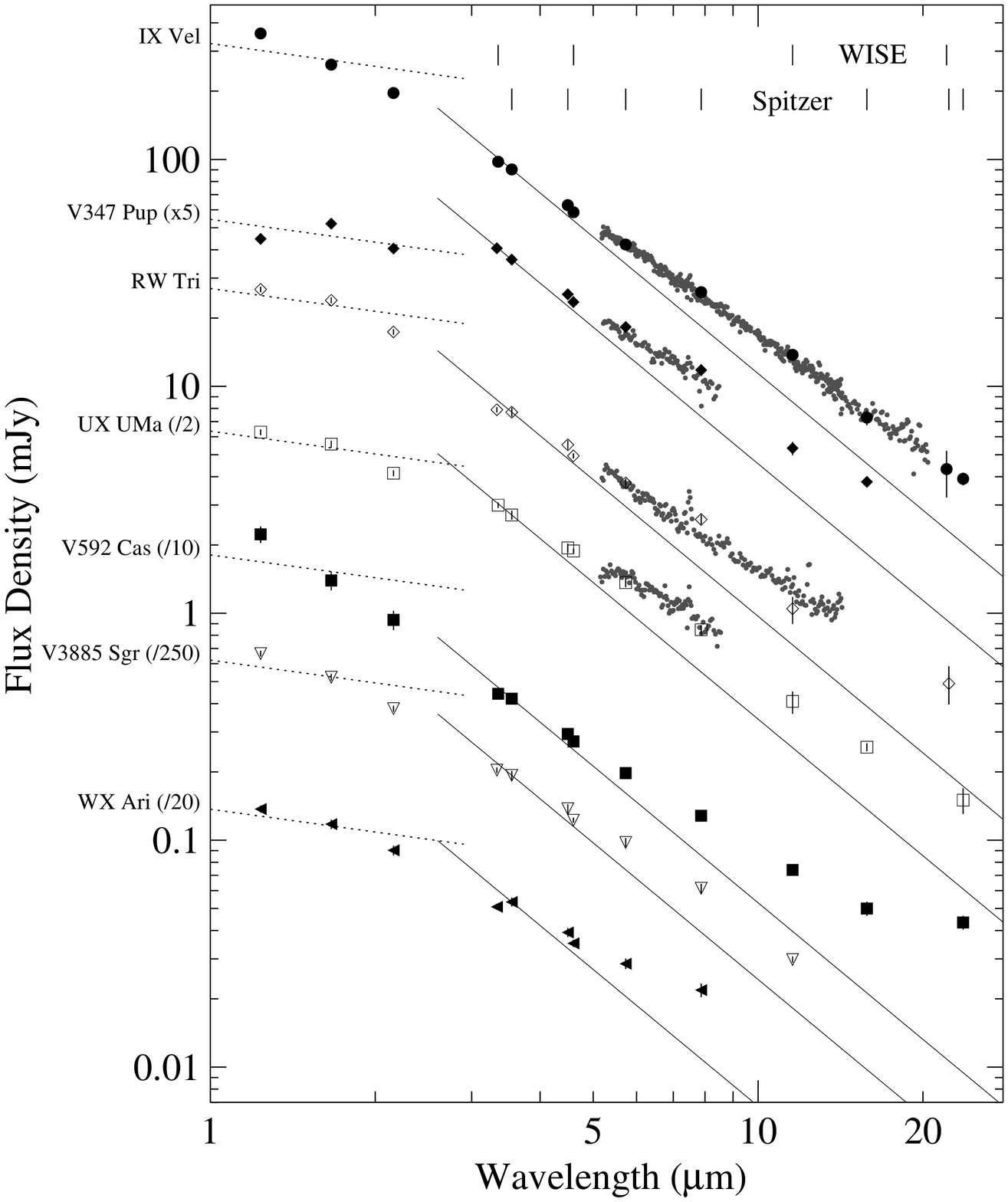}
\epsscale{1.0}
\caption{
As in Figure~\ref{f:seds02uncorrected}, but using the RJL-offset-corrected data where applicable (see Section~\ref{s:extractions} and individual target notes in Section~\ref{s:notes}) and the PSF-fit photometry value for the W3 point of V592~Cas (see Section~\ref{s:v592cas_data}).
\label{f:seds02}}
\end{figure*}

\begin{figure*}[tb]
\epsscale{0.80}
\plotone{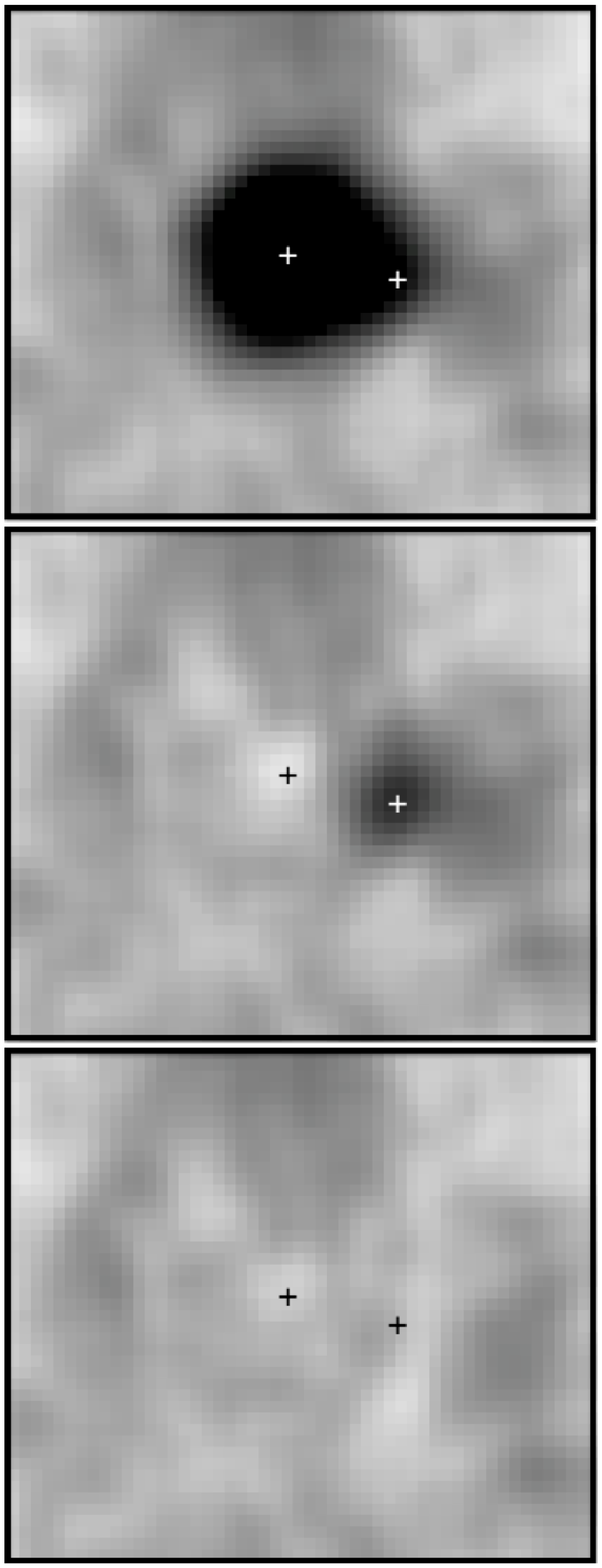}
\epsscale{1.0}
\caption{
{\it WISE} All Sky Release Atlas image of V592~Cas in the W3-band showing the overlapping profiles of V592~Cas and its bright neighbor (top panel).  The image is centered on the neighbor.  The middle panel shows the image after PSF-subtraction of the bright neighbor star, with only V592~Cas remaining.  The bottom panel shows the image after PSF-subtraction of both stars.  All panels have a width of 72\arcsec, plate scale of 1.375\arcsec\/ px$^{-1}$, and identical grayscale stretch.  Plus symbols in all of the panels mark the coordinates of the neighbor and V592 Cas from 2MASS and \citet{dws01}, respectively.
\label{f:v592cas_psfsub}}
\end{figure*}

The remaining targets, V347~Pup, RW~Tri, UX~UMa, V3885~Sgr, and WX~Ari 
(i.e., the targets shown in Figures~\ref{f:seds02uncorrected} and \ref{f:seds02}) 
have ``broken'' near-IR flux densities; that is, the J-to-H slope is different from the H-to-Ks slope (in the case of V347 Pup, the two slopes even have opposite signs).
In these five targets, there is generally better agreement with the characteristic disk spectrum prediction, especially with the J and H points.
On the other hand, the H-to-Ks slope is steeper than the $f_{\nu}\propto\lambda^{-1/3}$ prediction for all of the targets, 
suggesting that the transition to the Rayleigh-Jeans tail portion of the steady state accretion disk spectrum is occurring longward of H-band.

\citet{hoh04,hoh05} identified features from the secondary star in K-band spectra of most of a sample of two dozen CVs (primarily DNe) spanning a wide range of orbital period (1.9--11~hr).  They found that, unsurprisingly, the brighter, early type secondary stars in long period CVs are more easily detected, but also that the contribution from a bright accretion disk can make the identification and spectral typing of the secondary star difficult.  This situation is exacerbated among the NLs, in which the generally high mass transfer rates produce prominent disks that can dominate the SEDs over a wide range of wavelength.
\citet{dhillon00} found that in a sample of six bright NLs (including RW~Tri and VY~Scl), none show evidence of the secondary star in their K-band spectra. 
Nonetheless, we must consider that the secondary star could make a non-negligible contribution to the observed near-IR SED for some of our targets, which would produce disagreement with the \citet{FKR02} model accretion disk spectrum.  For example, while the secondary star in V592~Cas contributes only $\approx3$\% of the total light at H-band (see SED models in \citealt{hoard09} and Section~\ref{s:v592cas_model}), the secondary star contribution at H-band in IX~Vel is $\approx17$\% (see SED model in Section~\ref{s:ixvel_model}).
V347~Pup has a larger, hotter secondary star (M0.5; \citealt{thoroughgood05}) and smaller mass transfer rate ($6\times10^{-9}$~M$_{\odot}$~yr$^{-1}$; \citealt{puebla07}) than V592~Cas (M4.1, $1.5\times10^{-8}$~M$_{\odot}$~yr$^{-1}$; see below), so likely has more secondary star ``contamination'' of the disk spectrum in the near-IR.  Similarly, in UX~UMa, which has a relatively low mass transfer rate of $8\times10^{-9}$~M$_{\odot}$~yr$^{-1}$ (see below), the M2.6 secondary star contributes $\approx40$\% of the total light at H-band (see Section~\ref{s:uxuma_model}).



We fit the longer wavelength {\it Spitzer} and {\it WISE} data with a Rayleigh-Jeans tail spectrum, normalized to the IRAC-1 (3.55~$\mu$m) flux density point for each target.  
In all cases, the observed data show IR SEDs at $\lambda > 3.55$~$\mu$m with continuum slopes shallower than $f_{\nu}\propto\lambda^{-2}$; hence, their SEDs show an excess of flux density relative to the expectation of the standard steady state accretion disk theory.
This IR excess is apparent at wavelengths as short as $\approx4.5$~$\mu$m (IRAC-2, W2) and increases in strength at longer wavelengths.  
Clearly, there is a ``non-standard'' contribution to the IR SED in all of these CVs.

A pertinent question is whether the observed IR excess is due to a shortcoming in the standard model treatment or the presence of an additional system component.
The former option presumes that the spectrum of an accretion disk in the IR might not be well-represented by a sum of blackbodies.  There has been little work in detailed model atmosphere synthetic spectrum modelling of CV accretion disks in the IR (with most work to-date focussing on the ultraviolet spectral region). \citet{linnell07} and \citet{lhs07} have shown that from the optical out to $\approx3.5$~$\mu$m, the SED continuum shape returned by disk model atmosphere calculations still conforms to the expectation from the blackbody-based steady state theory described in \citet{FKR02}.  It remains to be seen whether the detailed physics of very cool gas contributing at even longer wavelengths could produce the observed SEDs.  We concentrate here on the latter option, in which the observed IR excess is due to the presence of a previously unconsidered physical component of the CV.

\begin{figure*}[tb]
\epsscale{0.75}
\plotone{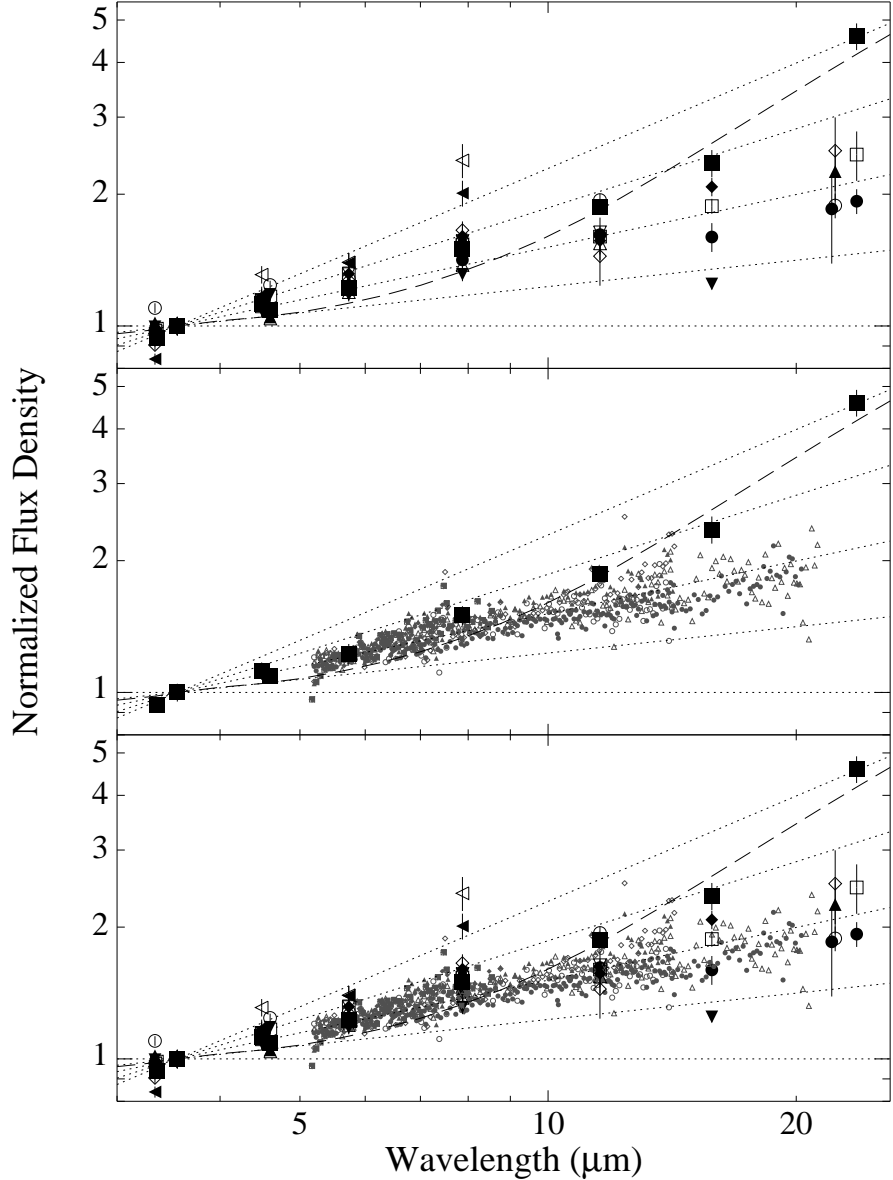}
\epsscale{1.0}
\caption{Normalized SEDs showing the data from Figures~\ref{f:seds01} and \ref{f:seds02} (with the same symbol coding), but normalized to the Rayleigh-Jeans model accretion disk SEDs (horizontal dotted line).  
An ordinate value of $y$ indicates that the observed flux density is a factor of $y$ larger than the value expected at that wavelength for the Rayleigh-Jeans disk model alone.
The top panel shows only the photometric data, the middle panel shows only the spectroscopic data, and the bottom panel shows all of the data.
For comparison, the V592~Cas photometric data (large filled squares) and system model (dashed line; see Section~\ref{s:v592cas_model}) are shown in all of the panels.
The dotted lines show normalized spectral profiles of $f_{\nu}\propto\lambda^{-\alpha}$, where $\alpha=2.0,1.9,1.8,1.7$, and 1.6, respectively from bottom to top.
\label{f:normdat}}
\end{figure*}

In Figure~\ref{f:normdat}, we have replotted the SED data from Figures~\ref{f:seds01} and \ref{f:seds02}, but have normalized the {\it Spitzer} and {\it WISE} data by dividing 
by the corresponding Rayleigh-Jeans tail accretion disk model shown in Figures~\ref{f:seds01} and \ref{f:seds02}, which was, in itself, normalized to the {\it Spitzer} IRAC-1 flux density.  The spectroscopic data only are then renormalized by multiplying by the ratio of the IRAC-3 (5.7~$\mu$m) flux density to the mean of each spectrum in the wavelength range 5.4--6~$\mu$m.  Thus, the horizontal dotted line at a constant normalized flux value of 1.0 in the figure shows the Rayleigh-Jeans disk model SED ($f_{\nu}\propto\lambda^{-\alpha}$, with $\alpha=2$), and the IR excess of each target relative to this model is readily apparent.  
Additional dotted lines in the figure show shallower spectral profiles with smaller values of the index $\alpha$. 

We note that there is no special physical significance to this (re)normalization process:\ it simply 
provides a means of easily comparing the IR excess in multiple CVs that have different distances and intrinsic luminosities.
The choice of the IRAC-1 point for normalizing the theoretical accretion disk spectrum is 
motivated primarily by the desire to choose a wavelength long enough that the ``standard'' CV components are on the Rayleigh-Jeans tail, but short enough that  any ``non-standard'' components make a negligibly small contribution.
As described above, there are obvious deviations in the JHKs bands from the Rayleigh-Jeans tail expectation of the steady state accretion disk theory for many of our targets.  
The shortest wavelength data available longward of Ks band
are from the W1 and IRAC-1 bands at $\approx3.5$~$\mu$m; the latter is preferred over the former because the IRAC-1 photometry has smaller uncertainties than W1.
Hindsight from the multi-component model fits 
confirms that the SEDs at 3.55~$\mu$m contain only a very small non-standard (IR excess) component but a large standard (accretion disk + secondary star) component 
(see Section~\ref{s:models}).

There are two features of note in Figure~\ref{f:normdat}:

First, V592~Cas has the most prominent long-wavelength ($\lambda\gtrsim15$~$\mu$m) IR excess of any of these targets.
At 7.9~$\mu$m (IRAC-4), only V442~Oph and WX~Ari have an IR excess larger than that of V592~Cas.
These two CVs are also both brighter than V592 Cas (relative to the disk model) at the shorter IRAC-2 (4.5~$\mu$m) and IRAC-3 (5.6~$\mu$m) wavelengths, although to a lesser extent than at IRAC-4 (7.9~$\mu$m), where they are the brightest CVs in our sample relative to the disk model.
However, IRAC-4 is the longest wavelength photometric point currently available for WX~Ari and V442~Oph, so it is uncertain if their IR excesses are significantly stronger than that of V592 Cas all the way to 24~$\mu$m.

Second, across the entire wavelength range of the data, but especially at $\lambda\lesssim10$~$\mu$m, the normalized photometry and spectroscopy for each target are remarkably similar.  
With the exceptions of V592~Cas, WX~Ari, and V442~Oph, the targets in our sample have IR SEDs consistent with a spectral index of $\alpha\approx1.8\pm0.1$ in the wavelength range $\Delta\lambda\approx3.5$--$25$~$\mu$m.
This implies that the spectral profile of the component producing the IR excess in these targets is similar.
The IR SEDs of WX~Ari and V442~Oph are consistent with a shallower spectral index of $\alpha\lesssim1.6$ out to their last available data points at 7.9~$\mu$m, while that of V592~Oph does not follow a simple power law over the entire wavelength range.

\subsection{Correlations with System Parameters}
\label{s:syspars}

\begin{figure*}[tb]
\epsscale{1.00}
\plotone{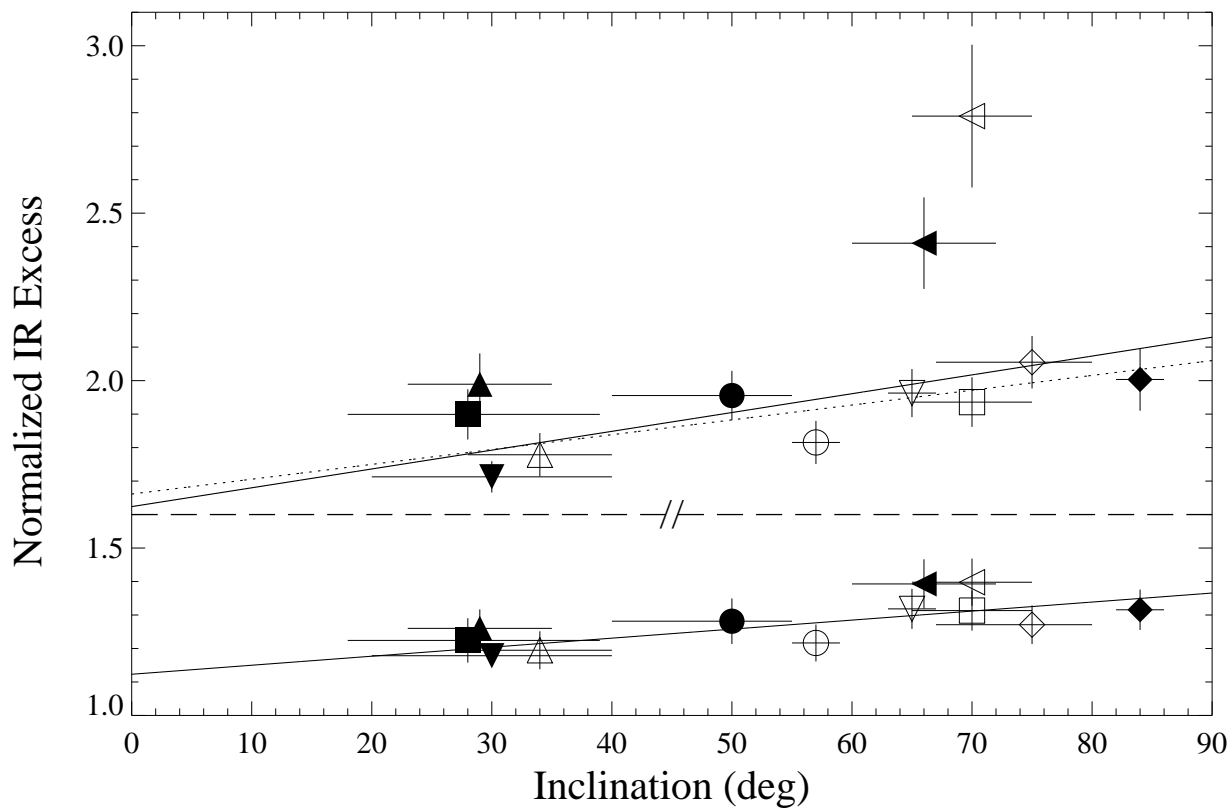}
\epsscale{1.0}
\caption{
Relation between system inclination (see Table~\ref{t:targets}) and normalized IR excess (from Figure~\ref{f:normdat}).  
The lower data show the IRAC-3 (5.7~$\mu$m) excess, while the upper data (separated by the broken dashed line) show the IRAC-4 (7.9~$\mu$m) excess (offset by +0.4 for clarity).
The solid lines are linear fits to each data set, weighted by the uncertainties on both parameters.
The dotted line is the linear fit to the 7.9~$\mu$m data after excluding the two outlying points.
Plot symbols are coded by target as in Figures~\ref{f:seds01}, \ref{f:seds02}, and \ref{f:normdat}.
\label{f:inclrat_incl}}
\end{figure*}

In Figure~\ref{f:inclrat_incl}, we show the relation between system inclination (from Table~\ref{t:targets}) and IR excess at 5.7 and 7.9~$\mu$m (the normalized IRAC-3 and IRAC-4 values from Figure~\ref{f:normdat}).  
Linear fits to each data set, weighted by the uncertainties of both parameters, reveal shallow slopes of increasing IR excess with increasing inclination.  
The fit parameters are listed in Table~\ref{t:corrs}.
There is little change in the linear fit results when the two outliers in the 7.9~$\mu$m excess data set, WX~Ari and V442~Oph, are excluded from both data sets.
As discussed above, these two targets -- along with V592~Cas -- have the strongest IR excesses in our sample.  V442~Oph and WX~Ari would benefit from additional long wavelength observations at $\lambda>8$~$\mu$m, since their IR excess appears to still be growing stronger at the longest available wavelength in their SEDs (IRAC-4).  
We also calculated the Spearman's rank correlation coefficient, $\rho$,
for these two cases (see Table~\ref{t:corrs}).
This allows us to make a more general assessment of whether these two parameters are related by a monotonic function of any form (not necessarily linear).
The results are consistent with those of the linear fits:\ 
there is a strong, statistically significant, positive correlation between inclination and IR excess at both wavelengths, regardless of whether or not the two outliers are excluded.  The probability of these correlations occurring by chance is small ($\leq10$\% in all cases, and $<5$\% in three of the four cases).

It is tempting to ascribe the shallow trend of increasing IR excess with increasing inclination to an increasing visibility of the relatively cool accretion disk rim (or, conversely, decreasing visibility of the hot inner disk) with increasing inclination.  
The observed change in normalized IR excess at 7.9~$\mu$m going from $i\approx30^{\circ}$ (when little accretion disk rim is directly visible) to $i\approx90^{\circ}$ (when only the disk rim is visible) amounts to an increase of $\approx20$\%.  
The SED of a model steady state accretion disk using the parameters for V592 Cas ($i=28^{\circ}$; see Sections~\ref{s:models} and \ref{s:v592cas_model}) is significantly fainter when recalculated for an inclination of $90^{\circ}$ (owing to the substantial decrease in overall projected emitting surface area).
The corresponding increase in normalized IR excess 
relative to 3.55~$\mu$m (IRAC-1) 
at 7.87~$\mu$m (IRAC-4) or 23.68~$\mu$m (MIPS-24)
is only $\approx4$\% or $\approx6.5$\%
(owing to the occultation of the hot inner disk and increased view of the cooler disk rim).
This argues against the accretion disk rim being the sole source of the observed IR excess\footnote{Not to mention that the standard model predicts temperatures of $\gtrsim6000$~K for the accretion disk rims of the NLs modelled in this work -- consistent with disk rim temperatures found by other modelling work \citep{linnell07,linnell08a,linnell08b,linnell10,linnell12} -- whereas the IR excess corresponds to much cooler thermal emission ($\lesssim1500$~K).  Cooler disk rim temperatures cannot be achieved within the constraint that the accretion disk must have a maximum radius no larger than the WD's Roche lobe (and, in fact, somewhat smaller still due to tidal truncation of the disk -- see the review in Section~2.5.5 of \citealt{warner03}, and references therein).}.

\begin{figure*}[tb]
\epsscale{1.00}
\plotone{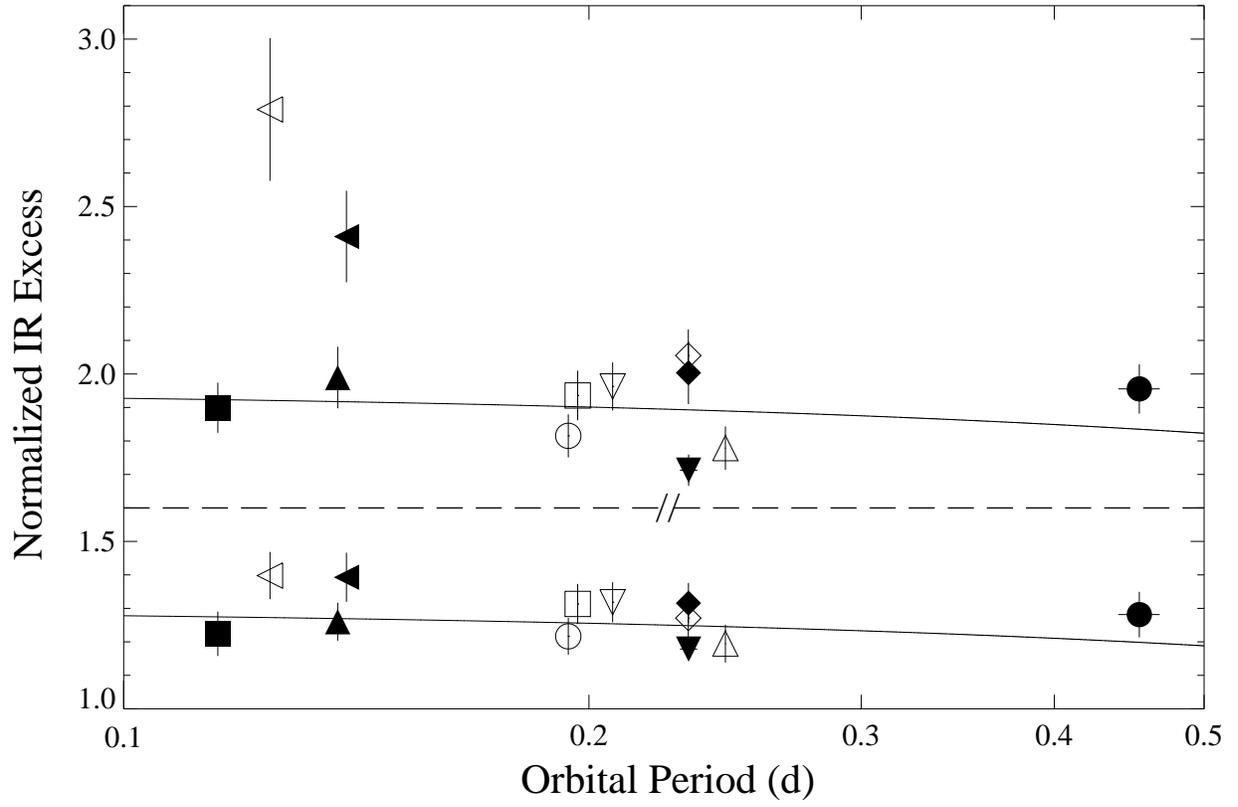}
\epsscale{1.0}
\caption{
As in Figure~\ref{f:inclrat_incl}, but for orbital period (see Table~\ref{t:targets}) and normalized IR excess.
\label{f:inclrat_porb}}
\end{figure*}

Figure~\ref{f:inclrat_porb} shows the comparable relation between orbital period and IR excess at 5.7 and 7.9~$\mu$m.  V442~Oph and WX~Ari are, again, outliers in the 7.9~$\mu$m data set (less so in the 5.7~$\mu$m data set).  
The linear fits to the complete data sets at both wavelengths, weighted by the uncertainties of both parameters, have negative slopes; however, they are consistent with zero within $\approx1\sigma$ (see Table~\ref{t:corrs}).  After excluding the two outliers, the fit slopes are significantly more shallow (consistent with zero within $\ll1\sigma$), implying that there is not a strong relationship between these two parameters.
The Spearman's rank coefficient shows a negative correlation in both cases.  The corresponding values of $\rho$ are smaller than for the correlations with inclination, and close to zero when the two outliers are excluded.  Additionally, there is a high probability that the correlations at both wavelengths (both with and without the outliers) could have occurred by chance.
Thus, there appears to be no statistically significant correlation between orbital period and IR excess.
This implies that the strength of the IR excess is not strongly linked to differences in mass transfer rate caused by secular evolution of the CVs (keeping in mind that the NLs generally have high mass transfer rates compared to other CVs, so a link between orbital period and/or mass transfer rate and the IR excess could be a higher order effect that is not apparent here).

\subsection{Spectral Energy Distribution Models}
\label{s:models}

In this section, we present physically realistic, multi-component models calculated as in \citet{hoard09} to reproduce the IR SEDs of several representative members of our target sample.  The similarity of the SEDs (cf.\ Figure~\ref{f:normdat}), parameter degeneracies of the accretion disk and circumbinary dust disk model components (see below), and the fact that almost all of these targets have well-determined physical parameters or complete system models already in the literature, suggests that there would be a limited scientific return on reproducing the SEDs of all of the CVs in our sample.  Rather, we can infer from a few representative models that targets with similar SEDs will have similar models and similar components that fit their IR excess.
Parameters used for the individual models discussed below are listed in Table~\ref{t:models1}; published system parameters for the other targets are listed in Table~\ref{t:models2}.  The following information applies to all of the models:

\begin{enumerate}
\item{The WD is represented by a blackbody curve of the desired temperature ($T_{\rm wd}$), normalized to the radius ($R_{\rm wd}$) implied by the mass of the WD ($M_{\rm wd}$).}

\item{The secondary star is the semi-empirical template with spectral type (Sp[2]), temperature ($T_{2}$), and mass ($M_{2}$) appropriate for the target's orbital period from \citet{knigge11}, who performed a comprehensive analysis of the physical properties of the secondary stars across the entire population of CVs.
The templates provided in \citet{knigge11} include only the L (3.5~$\mu$m), L$'$ (3.8~$\mu$m), and M (4.80~$\mu$m) bands in the IR.  Consequently, we have calculated appropriate synthetic photometry for the {\it WISE} and {\it Spitzer} (IRAC, MIPS-24) bands to substitute for the L, L$'$, M band values in the published templates.  These new photometric values were calculated as described in \citet{knigge06} and \citet{knigge11}, but using the BT-Settl model atmospheres\footnote{See \url{http://phoenix.ens-lyon.fr/simulator/index.faces}.} \citep{agl03,aah07,ahf12} with the grid of secondary star physical parameters from Table~6 in \citet{knigge11}.
}

\item{The accretion disk (ACD) component assumes an optically thick steady state disk composed of concentric rings emitting as blackbodies, following the standard model prescription in \citet{FKR02}. The accretion disk component is parameterized by inner and outer radii ($R_{\rm acd, in}$ and $R_{\rm acd, out}$), vertical semi-height at the inner radius ($h_{\rm acd}$), and a mass transfer rate from the secondary star (\.{M}).  The accretion disk is composed of 100 concentric rings, each of which contains 360 azimuthal sections.

The temperature of each disk ring is determined from the standard model radial temperature profile ($T\propto r^{-\gamma}$ where $r$ is radial distance from the disk center and $\gamma=0.75$; \citealt{FKR02}).  The disk is flared such that the height of the disk for radii larger than $R_{\rm crit} = [(7 - 8\gamma_{\rm out})/(6 - 8\gamma_{\rm out})]^2 \, R_{\rm wd}$ increases as $r^{9/8}$ \citep{warner03}.  In this case, $\gamma_{\rm out} < 0.75$ provides a shallower temperature gradient in the outer disk corresponding to irradiation of the inner face of the flared disk by the inner disk and WD \citep{orosz03}; we use $\gamma_{\rm out} = 0.70$ \citep{hoard09}.  The vertical outer edge of the disk, which does not ``see'' the WD and inner accretion disk, is assumed to have a non-irradiated temperature appropriate for the standard model radial temperature profile.  
Depending on the system inclination and disk geometry, the
model accounts for self-occultation of the inner disk by excluding the flux contribution of regions that are blocked from view by the flared disk.

A wavelength- and temperature-dependent correction for limb darkening of the accretion disk is applied to each azimuthal section using linear limb-darkening coefficients interpolated over the grid provided by \citet{vanhamme93}. In the case of a flat disk, this would be a relatively simple procedure, as the disk face everywhere presents the same viewing angle (i.e., angle between the line of sight and a normal to the disk face), so only the disk edge is limb-darkened across a range of viewing angles. However, in the case of the flared disk used here, which presents different viewing angles for different parts of the disk, limb darkening is calculated individually for each azimuthal section in each ring. The primary effect of limb darkening is to make the short wavelength (UV--optical) end of the SED fainter relative to the long wavelength (IR) end.

As found by \citet{orosz03} and other studies of irradiation in model accretion disk SEDs (e.g., \citealt{wade88}), we found that irradiation (like limb darkening) primarily affects the short wavelength end of the SED. However, irradiation (which effectively raises the temperature at a given radius over the non-irradiated case) tends to make the short wavelength end of the SED brighter relative to the long wavelength end, which somewhat counteracts the effect of limb darkening.}

\item{The circumbinary dust disk (CBD) component is calculated as described in \citet{hoard07} and \citet{hoard09}, under the assumption that the disk is optically thin and composed of spherical grains with 
characteristic
radius of $r_{\rm grain}=1$~$\mu$m and density of $\rho_{\rm grain}=3$~g~cm$^{-3}$ that (re)radiate as blackbodies. 
The inner edge of the disk is fixed at the tidal truncation limit ($R_{\rm cbd, in}\approx1.5$ times the binary separation), and the outer edge ($R_{\rm cbd, out}$) is fixed at the distance at which the temperature of the dust merges with the ambient interstellar medium at a temperature of 20~K.  The input parameters for the CBD component are only the inner edge temperature, $T_{\rm cbd, in}$ (keeping in mind that silicate dust will sublimate at temperatures higher than $\approx1500$--$2000$~K; e.g., \citealt{pollack94,kobayashi11}) and the exponent of the radial temperature profile ($\gamma=0.75$).  

As noted in \citet{hoard07} and \citet{hoard09}, the degeneracies among parameters in the CBD model mean that any given model is representative of a class of similar possible solutions, but is not necessarily a unique solution.  Of particular note in the context of the work presented here is the degeneracy between $r_{\rm grain}$ and total dust mass ($M_{\rm cbd,total}$) for constant $\rho_{\rm grain}$.
The total dust mass values for the models listed in Table~\ref{t:models1} were calculated assuming a characteristic dust grain radius of 1~$\mu$m.  
For constant grain mass density, the total dust mass required to produce an identical CBD SED component scales with $r_{\rm grain}$.

The use of a characteristic grain radius is a simplifying assumption that can be interpreted as a representative (or approximate mean) value among a distribution of sizes, for those grains that predominantly contribute to the observed SED (e.g., see discussion in the appendix of \citealt{jura09}). 
On a grain-by-grain basis, the emitted spectrum of a dust grain in temperature equilibrium is a blackbody modified by a function that depends on wavelength (as well as temperature, grain size and composition; e.g., \citealt{dss80,draine81}).
This function is approximately a constant (i.e., independent of wavelength) for $\lambda\lesssim\pi r_{\rm grain}$ (i.e., $\sim3$--$30$~$\mu$m for 1--10~$\mu$m grains), and varies as $\lambda^{-1}$--$\lambda^{-2}$ at wavelengths much larger than the grain size, $\lambda\gtrsim5\pi r_{\rm grain}$ (i.e., $\sim16$--$160$~$\mu$m for 1--10~$\mu$m grains; e.g., \citealt{cbm04}).  
Nonetheless, in the case of dust disks around single WDs, blackbody SED models utilizing an appropriate radial temperature profile adequately reproduce the broad-band IR photometric observations (e.g., \citealt{jura03,farihi11}), under the assumption that the grains are predominantly up to $\sim$ a few $\mu$m in size (based on the presence of the 10~$\mu$m silicate emission feature -- see additional discussion of this in Section \ref{s:missing}) with a characteristic size of $\sim1$~$\mu$m \citep{jura09}.  
If the distribution of sizes is dominated by grains that are much smaller ($\lesssim0.1$~$\mu$m) or much larger ($\gtrsim100$~$\mu$m), then the resultant near- to mid-IR spectrum could deviate significantly from the expectation of simple blackbody radiation.
}  

\item{For a given target, all model components are normalized to the estimated distance to the target ($d$) and are assumed to share a single inclination ($i$). The values of $d$ and $i$ are taken from best estimates in the literature (see Table \ref{t:targets}) and, in the interests of improving the degree to which the available data can constrain the models, are not allowed to vary freely.}
\end{enumerate}

The f$_{{\rm acd,}\lambda}$ and f$_{{\rm cbd,}\lambda}$ values listed in Table~\ref{t:models1} are the fractions of the total modelled system light contributed by the accretion disk and circumbinary disk components, respectively, at the indicated wavelengths.  The contribution from the WD is negligible at the tabulated IR wavelengths, so the secondary star contribution can be inferred by subtracting the sum of the tabulated fractions from 100\%.

\subsubsection{Model for V592 Cas}
\label{s:v592cas_model}

\begin{figure*}[tb]
\epsscale{1.00}
\plotone{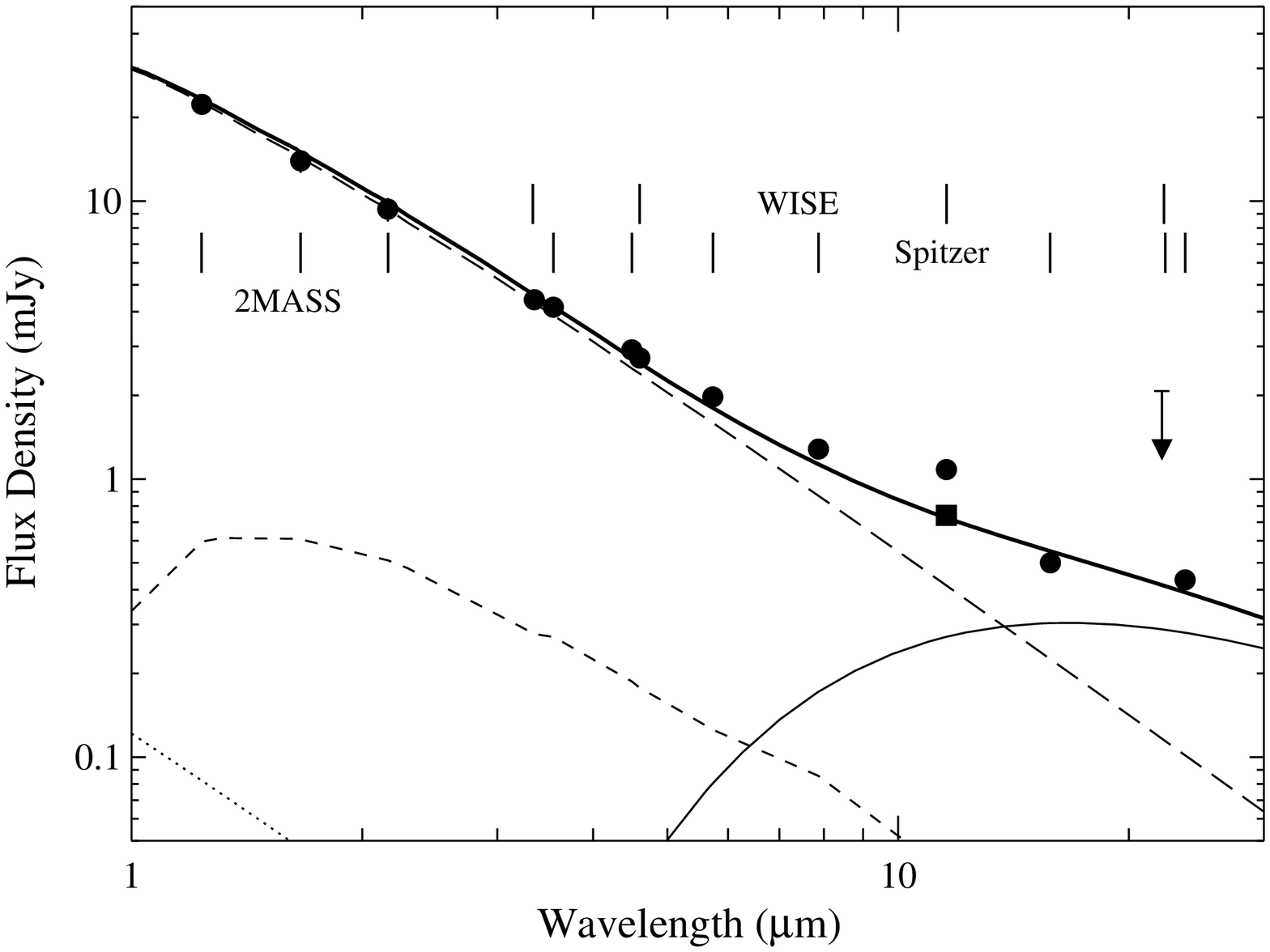}
\epsscale{1.0}
\caption{
IR SED of V592~Cas 
using the 2MASS (dereddened), {\it WISE}, and re-extracted {\it Spitzer} photometry 
(circles), along with the multi-component system model from \citet{hoard09} and this work.
The total system model (thick solid line) is composed of a WD (dotted line), M4.1 secondary star (short dashed line; semi-empirical template from \citealt{knigge11} with additional IR points), 
limb-darkened, flared and irradiated, steady state accretion disk (long dashed line), and circumbinary dust disk (thin solid line). 
The isophotal wavelengths of the 2MASS, {\it WISE}, and {\it Spitzer} (IRAC, IRSPUI, MIPS-24) bands are indicated.  
The {\it WISE} W4 point is a non-detection upper limit.  The single square point is our PSF-fit photometry for the {\it WISE} W3 band (see Section~\ref{s:v592cas_data}).
\label{f:v592cas_model}}
\end{figure*}

Figure~\ref{f:v592cas_model} shows our model for V592~Cas from \citet{hoard09} overlaid on the re-extracted {\it Spitzer} photometry discussed here (along with 2MASS and {\it WISE}).  The only difference from the \citet{hoard09} treatment is that we have used the updated secondary star model from \citet{knigge11} appropriate for the orbital period of V592~Cas 
(M4.1; see Table~\ref{t:models1})
-- this requires a slightly larger (by $\approx4$\%) distance than used in \citet{hoard09} to scale the model to the data.  
There is no significant disagreement with the model due to the slightly different photometry yielded by the final {\it Spitzer} cryogenic flux calibration.  
Unfortunately, the {\it Herschel} upper limits at 70 and 160~$\mu$m for V592~Cas, $\le1.1$~mJy and $\le4.5$~mJy, respectively \citep{harrison13}, are not helpful in constraining the presence and characteristics of dust.  They are more than an order of magnitude brighter than the extrapolation of the V592~Cas SED model, which gives a total predicted system flux density at $\gtrsim70$~$\mu$m of $<0.1$~mJy (dominated by the CBD component).

\subsubsection{Model for IX Vel}
\label{s:ixvel_model}

\begin{figure*}[tb]
\epsscale{1.00}
\plotone{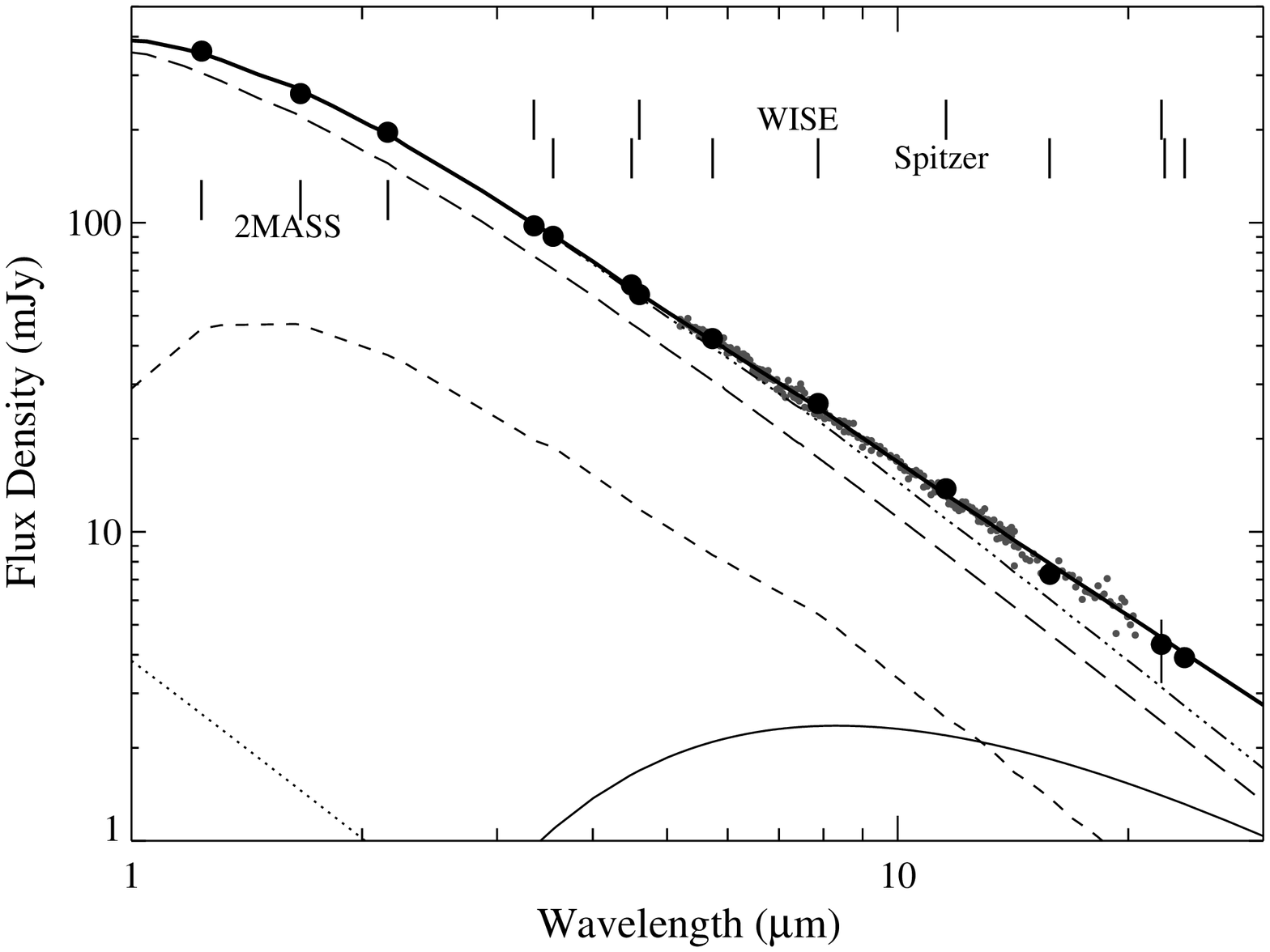}
\epsscale{1.0}
\caption{
IR SED of IX~Vel using the 2MASS, {\it WISE}, {\it Spitzer} photometry (circles), and {\it Spitzer} spectrum (small circles), along with our multi-component system model.
The total system model (thick solid line) is composed of a WD (dotted line), M2.6 secondary star (short dashed line; semi-empirical template from \citealt{knigge11} with additional IR points), limb-darkened, flared and irradiated, steady state accretion disk (long dashed line), and circumbinary dust disk (thin solid line). 
The system model utilizes the parameters of \citet{linnell07} (see Tables~\ref{t:targets} and \ref{t:models1}).
The dot-dot-dot-dash line is not an additional model component; instead, it shows the model without the contribution of the circumbinary dust disk.
The isophotal wavelengths of the 2MASS, {\it WISE}, and {\it Spitzer} (IRAC, IRSPUI, MIPS-24) bands are indicated.  
\label{f:ixvel_modelB}}
\end{figure*}

\citet{linnell07} calculated a model fit to the UV--optical SED of IX~Vel with the BINSYN multi-component model atmosphere/synthetic spectrum code \citep{linnell96,linnell08b,linnell12}.
We utilize their system parameters as starting points (see Table~\ref{t:models1}) to construct our IR SED model.  
Figure~\ref{f:ixvel_modelB} shows our model for IX~Vel, composed of WD, secondary star, accretion disk, and CBD components.  

\begin{figure*}[tb]
\epsscale{1.00}
\plotone{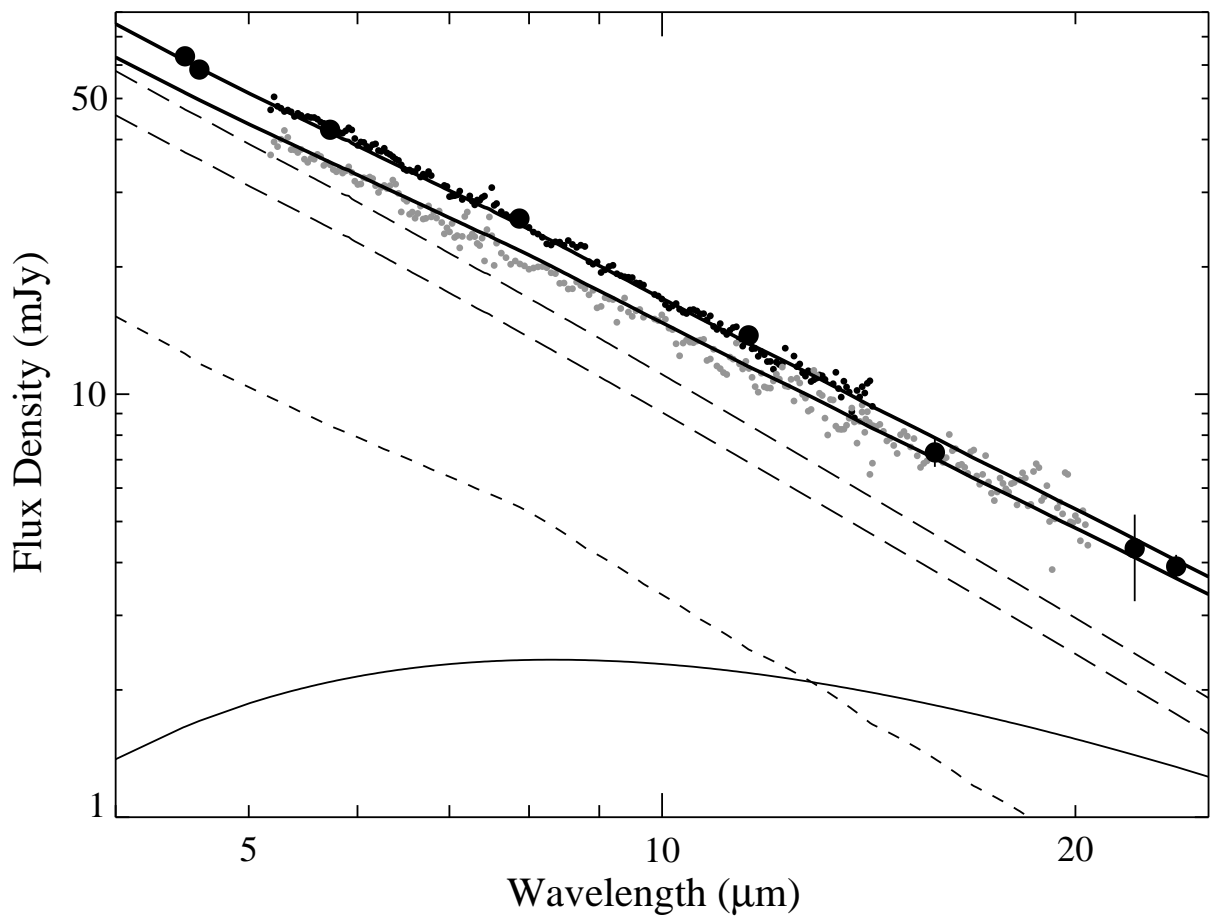}
\epsscale{1.0}
\caption{
Comparison of accretion disk models for IX~Vel.  This is a close-up of the IR region of the model shown in Figure~\ref{f:ixvel_modelB}, with the Cycle-2 {\it Spitzer} data (black points) and corresponding secondary star (short dashed line), CBD (thin solid line), accretion disk (upper long dashed line), and total system (upper thick solid line) model SEDs.
The uncorrected Cycle-5 IRS spectrum (small gray points) is plotted along with a fainter accretion disk model component (lower long dashed line) that yields a total system model (lower thick solid line) in agreement with the Cycle-5 spectrum.  See Section~\ref{s:ixvel_model} for additional discussion.
\label{f:ixvel_modelD}}
\end{figure*}

In Section~\ref{s:ixvel_data}, we applied a correction to the Cycle-5 IRS spectrum of IX~Vel to bring it into agreement with the brighter Cycle-2 IRS spectrum.  We now demonstrate that the assumed spectral shape of this correction (i.e., a Rayleigh-Jeans-like $f_{\nu}\propto\lambda^{-2}$), which corresponds to the assumption that the overall brightness change is dominated by a change in the accretion disk modulated by \.{M}, is justified.
Figure~\ref{f:ixvel_modelD} shows a close-up of the IR region of the IX Vel SED and model shown in Figure~\ref{f:ixvel_modelB}.  The Cycle-2 {\it Spitzer} data are plotted in black, and the uncorrected (fainter) Cycle-5 IRS spectrum is plotted in gray.  The secondary star (short dashed line) and CBD (thin solid line) components are unchanged from the model shown in Figure~\ref{f:ixvel_modelB} (and described above), as are the brighter accretion disk (upper long dashed line) and system (upper thick solid line) models.  The fainter accretion disk model component (lower long dashed line) has \.{M}~$=3.4\times10^{-9}$~M$_{\odot}$~yr$^{-1}$, which is $\approx50$\% of the bright accretion disk value.  The corresponding faint system model (lower thick solid line) agrees well with the uncorrected Cycle-5 data.  

The difference between the two accretion disk components differs from a pure $\lambda^{-2}$ dependence by less than 4\% at all wavelengths in the range 5--20~$\mu$m.
Hence, the functional form of the correction is reasonable in the context of the assumption stated above.
In any case, the outcome of applying the RJL correction here is primarily cosmetic -- none of the IR excess detections or modelling results depend solely on data that were corrected in this manner.
Incidentally, we note that a decrease in the accretion rate by a factor of 2 corresponds to a decrease in the total disk luminosity by a factor of 2 \citep{FKR02}; comparing the two accretion disk models for IX~Vel on a wavelength-by-wavelength basis shows that the corresponding change in the total system brightness would be $\Delta R\approx0.5$~mag and $\Delta K\approx0.3$~mag.

\subsubsection{Model for UX UMa}
\label{s:uxuma_model}

\begin{figure*}[tb]
\epsscale{1.00}
\plotone{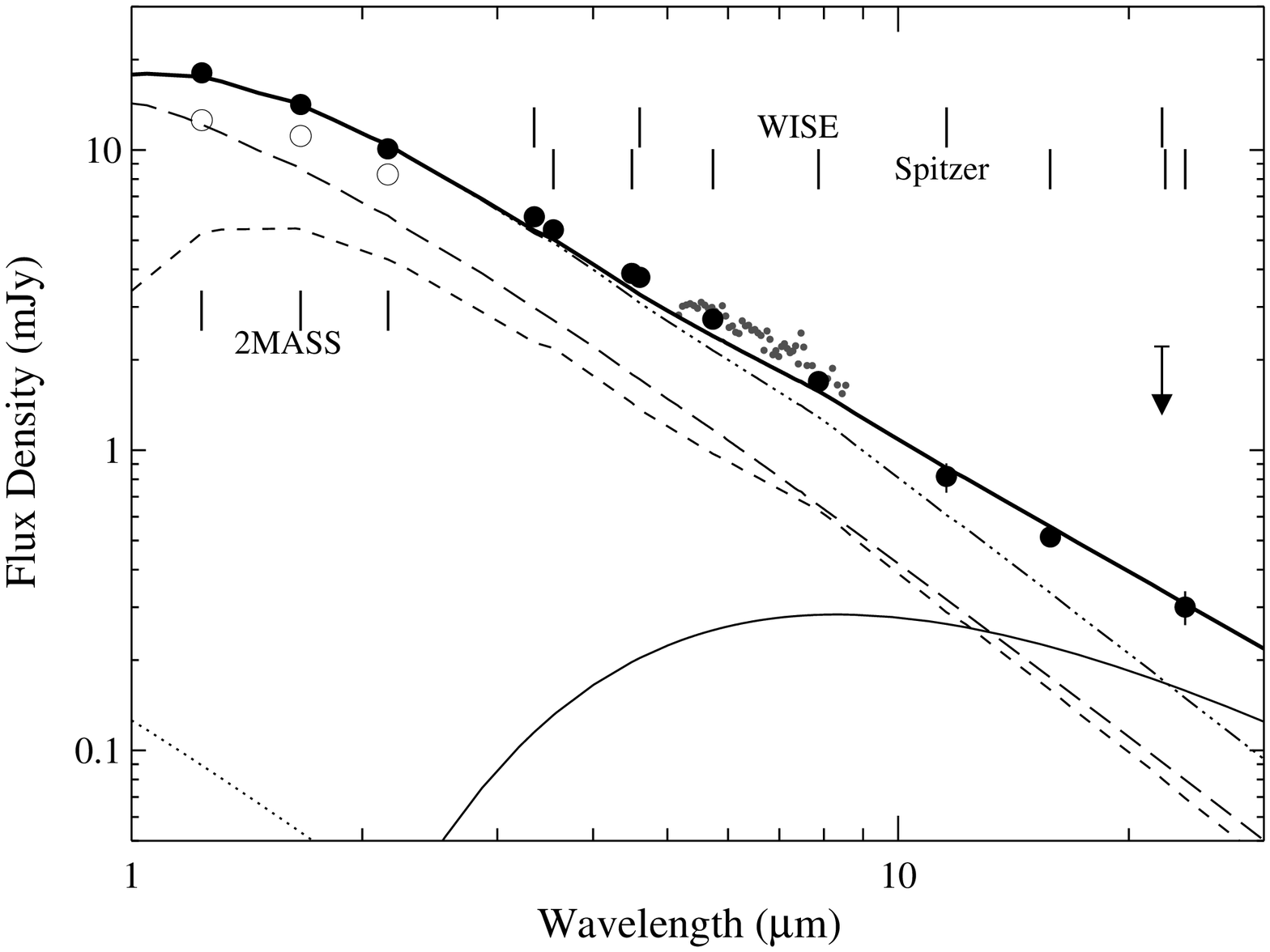}
\epsscale{1.0}
\caption{
IR SED of UX~UMa using the 2MASS, {\it WISE}, {\it Spitzer} photometry (circles), and {\it Spitzer} spectrum (small circles), along with our multi-component system model.
The total system model (thick solid line) is composed of a WD (dotted line), M2.6 secondary star (short dashed line; semi-empirical template from \citealt{knigge11} with additional IR points), limb-darkened, flared and irradiated, steady state accretion disk (long dashed line), and circumbinary dust disk (thin solid line) -- see model parameters in Table~\ref{t:models1}. 
An offset has been applied to the 2MASS points, as described in Section~\ref{s:uxuma_model}; the original 2MASS photometry is shown as unfilled circles.
The dot-dot-dot-dash line is not an additional model component; instead, it shows the model without the contribution of the circumbinary dust disk.
The isophotal wavelengths of the 2MASS, {\it WISE}, and {\it Spitzer} (IRAC, IRSPUI, MIPS-24) bands are indicated.  
\label{f:uxuma_modelB}}
\end{figure*}

\citet{linnell08a} calculated a model fit to the UV--optical SED of UX~UMa, using the BINSYN code as for IX~Vel.  However, in the case of UX~UMa, they were not able to completely resolve a degeneracy between mass transfer rate and distance, finding a range of equally acceptable solutions for $d=250$--$345$~pc and \.{M}~$=5$--$10\times10^{-9}$ M$_{\odot}$ yr$^{-1}$.
We started our modeling process using their suggested compromise values of $d=312$~pc and \.{M}~$=8\times10^{-9}$~M$_{\odot}$~yr$^{-1}$, as well as the other system parameters derived in their study (see Table~\ref{t:models1}).  However, it quickly became apparent that there was a flux density offset between the 2MASS and {\it Spitzer\/}+{\it WISE} data, with the former being fainter than expected compared to the latter.
The AAVSO long-term light curve of UX~UMa shows that there was no significant difference in mean brightness during the 2MASS observations compared to the later {\it Spitzer} and {\it WISE} observations.  However, UX~UMa -- like other CVs -- does display stochastic brightness variations, on timescales of days, with an amplitude of 0.2--0.3 mag (in V).  When we utilize only the {\it Spitzer} and {\it WISE} data, we obtain a good model for the IR SED using values of $d=290$ pc and \.{M}~$=7\times10^{-9}$~M$_{\odot}$~yr$^{-1}$, 
consistent with the range found by \citet{linnell08a}; however, the degeneracy between distance and \.{M} persists so this should be regarded as a representative, rather than unique, solution.
Adding a RJL offset to the 2MASS photometry, with an initial value of +5.5 mJy at J-band, brings it into conformity with the model.  We note that, in this case, the appropriateness of applying this correction is less certain, since the peak of the secondary star component makes a non-negligible ($\approx30$\%) and non-RJL contribution to the total system brightness in the 2MASS bands.
We plot both the original and corrected 2MASS data in Figure~\ref{f:uxuma_modelB}, which shows our SED model.

\subsubsection{Model for RW Sex}

\begin{figure*}[tb]
\epsscale{1.00}
\plotone{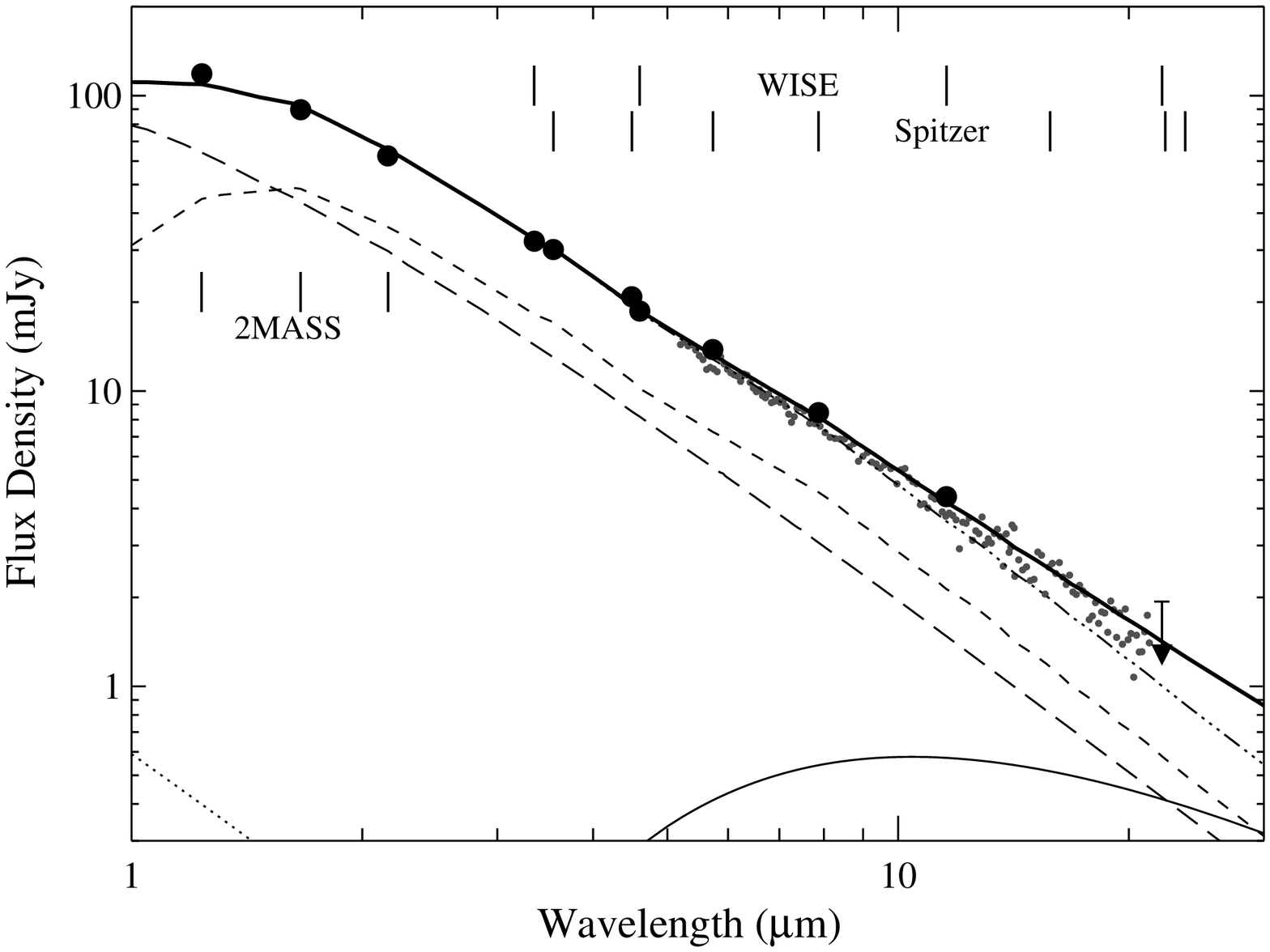}
\epsscale{1.0}
\caption{
IR SED of RW~Sex using the 2MASS, {\it WISE}, {\it Spitzer} photometry (circles), and {\it Spitzer} spectrum (small circles), along with our multi-component system model.
The total system model (thick solid line) is composed of a WD (dotted line), M1 secondary star (short dashed line; semi-empirical template from \citealt{knigge11} with additional IR points), limb-darkened, flared and irradiated, steady state accretion disk (long dashed line), and circumbinary dust disk (thin solid line) -- see model parameters in Table~\ref{t:models1}. 
The dot-dot-dot-dash line is not an additional model component; instead, it shows the model without the contribution of the circumbinary dust disk.
The isophotal wavelengths of the 2MASS, {\it WISE}, and {\it Spitzer} (IRAC, IRSPUI, MIPS-24) bands are indicated.  
\label{f:rwsex_modelB}}
\end{figure*}

We started our modeling process for RW Sex using the favored BINSYN model parameters from \citet{linnell10} (see Table~\ref{t:models1}).
The orbital period of this NL (353 min) is slightly longer than the longest orbital period for which \citet{knigge11} provide a semi-empirical secondary star template (336 min).  To compensate for this, we used the longest period secondary star template from \citet{knigge11}, scaled up in brightness by 15\%, which is the approximate difference between the longest period secondary star template and the secondary star template corresponding to an orbital period that is shorter by the same amount as the orbital period of RW~Sex is longer.
The resultant secondary star model component is slightly brighter than the accretion disk component obtained using the size and mass transfer rate parameters from \citet{linnell10}.
Figure~\ref{f:rwsex_modelB} shows the complete SED model for RW~Sex, utilizing the scaled \citet{knigge11} secondary star template and a CBD component to model the IR excess apparent at wavelengths longer than $\approx8$~$\mu$m.

\section{Discussion}
\label{s:discussion}

The IR SEDs of the eleven NLs presented here are largely similar to that of V592~Cas and all show an IR excess.  This suggests that the IR excess in all cases could have a similar origin.
We have demonstrated here (and in \citealt{hoard09}) that physically realistic, multi-component models utilizing emission from circumbinary dust disks can reproduce the observed SEDs.
At the same time, several of the targets observed here are believed to have significant accretion disk wind outflows, which can even exceed the WD escape velocity (e.g., see \citealt{hss95}, \citealt{pr95}, and Sections \ref{s:app_v592cas} and \ref{s:app_rwsex}), and could provide a suitable environment for the generation of bremsstrahlung (free-free emission) that also might produce an IR excess.
\citet{harrison13} investigated the mid- to far-IR properties of a sample of primarily DNe and magnetic CVs, with only the NL V592~Cas in common with our study.  They suggest that bremsstrahlung contributes significantly to the observed IR excesses in V592~Cas and the archetype DN SS~Cygni, with a possible additional contribution from dust only in V592~Cas.
Thus, we should devote some consideration to the relative merits of emission from circumbinary dust vs.\ bremsstrahlung as the explanation for the IR excesses in these objects.

It should first be noted, however, that there is a potentially serious drawback to the modelling approach used by \citet{harrison13}.  
To reproduce the optical--IR SEDs of their sample of CVs (including V592 Cas), \citet{harrison13} utilized a self-described ``simple'' model intended to avoid the difficulty in constraining a multi-component model (as in \citealt{hoard09} and this work).
They simply utilized two blackbodies, one representing all of the hot, optically thick regions in the CV (i.e., WD, boundary layer, accretion disk, etc.), and a second blackbody representing the secondary star.  Their justification for this modeling approach is that since all of the components are on the Rayleigh-Jeans tail in the IR, they simply sum together as a power law.  

\begin{figure*}[tb]
\epsscale{1.00}
\plotone{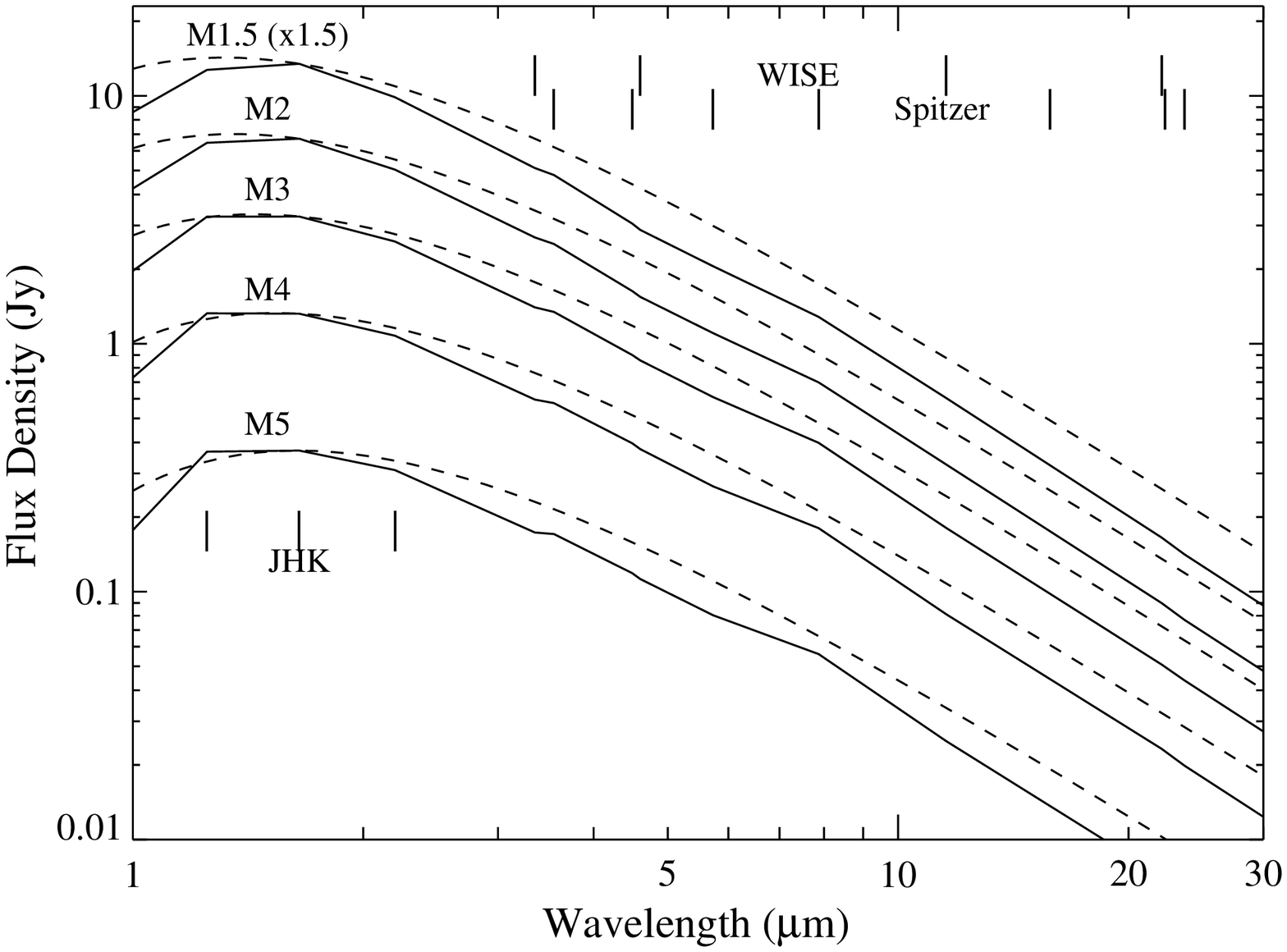}
\epsscale{1.0}
\caption{
Comparison of the \citet{knigge11} semi-empirical secondary star templates with additional IR points used here (solid curves) with blackbodies (dashed curves).
The secondary star templates are (top to bottom): 
M1.5 ($P_{\rm orb}=5.5$~hr, T$_2=3881$~K), 
M2   ($P_{\rm orb}=5.2$~hr, T$_2=3758$~K),
M3   ($P_{\rm orb}=4.3$~hr, T$_2=3532$~K),
M4   ($P_{\rm orb}=3.4$~hr, T$_2=3330$~K), and
M5   ($P_{\rm orb}=1.9$~hr, T$_2=3112$~K).
The flux densities correspond to a distance of 10~pc; the SED for spectral type M1.5 was scaled upward by a factor of 1.5 for clarity.
The blackbody functions were calculated for the same temperatures as the corresponding secondary star templates, and scaled to match the H-band point.
The isophotal wavelengths of the J, H, K, {\it WISE}, and {\it Spitzer} (IRAC, IRSPUI, MIPS-24) bands are indicated.  
\label{f:msed}}
\end{figure*}

The problem arises from the fact that cool (M type) stars deviate from having Rayleigh-Jeans spectra 
in the mid-IR (IRAC bands and longer wavelengths) 
due to molecular pollution in their photospheres \citep{patten06}.  In fact, the observed Ks$-$[MIPS-24] color index for cool stars ($T_{\rm eff}=2000$--$4000$~K) is systematically and significantly smaller than that of a blackbody at the corresponding temperature (see Figure~2 in \citealt{gautier07}).  Thus, if the star and the blackbody have the same Ks magnitude, then the blackbody will be brighter than the star at 24~$\mu$m. 
This effect is demonstrated in Figure~\ref{f:msed}, in which we have plotted several of the \cite{knigge11} secondary star templates with our added IR points spanning spectral types of M1.5--M5 (corresponding to CV orbital periods of 5.5--1.9~hr, respectively).  We have also plotted a corresponding blackbody function with the same temperature as each secondary star template, scaled to match the H-band flux density of the corresponding template.  
The blackbodies overestimate the SEDs of the secondary stars by 30--60\% in the {\it WISE} and {\it Spitzer} bands.
In V592~Cas, the ultimate effect of this is not severe, 
since the overall SED is dominated by the accretion disk at all wavelengths (e.g., the blackbody is $\approx30$\% brighter than the secondary star template at 10~$\mu$m, but this difference corresponds to $\approx2$\% of the total system brightness at that wavelength).  However, 
in systems for which the secondary star makes a larger relative contribution, using a blackbody to represent the secondary star could yield a false-negative for the presence of an IR excess.
For example, the blackbody ``secondary stars'' for UX~UMa and RW~Sex are $\approx25$\% and $\approx40$\% brighter than the \citet{knigge11} templates in the IRAC-4 (7.8~$\mu$m) band.  Since the secondary star contributes $\approx40$\% and $\approx50$\% of the total flux for UX~UMa and RW~Sex, respectively, in the IRAC-4 band, using a blackbody to represent the secondary star would overestimate the corresponding flux density by 10--20\%.  This is comparable to the strength of the IR excess component at the IRAC-4 band in these two systems (see Table~\ref{t:models1}), resulting in a false negative for the presence of an IR excess.  The negative results reported for the presence of IR excess in the CVs modeled by \citet{harrison13} using blackbody secondary stars (which dominate their model SEDs in the IR) could have been similarly affected and should be treated with caution.

\subsection{Dust or Bremsstrahlung?}
\label{s:brems}

We now compare the principal advantages and disadvantages of the dust and bremsstrahlung mechanisms as explanations for the observed IR excess in CVs.
Dust is 
known to be associated with some phases of CV activity, such as classical nova outbursts \citep{bode82,evans97,evans01,eg12} and tremendous outburst amplitude DNe \citep{cwh06}.
In addition, \citet{bisikalo09,bisikalo11} and \citet{skb07,sbk09} have performed detailed numerical simulations of the mass transfer process in CVs, and find that a significant fraction of the mass transferred from the secondary star into the WD Roche lobe is ejected from the inner binary through the outer Lagrange point (L$_{3}$).  This material spreads into circumbinary space around the CV (see Figure~5 in \citealt{bisikalo09}), and could provide the building blocks for dust formation (or even already-complete dust grains transferred from their formation site in the outer layers of 
a cool secondary star).  
Along with other possible delivery mechanisms for material into circumbinary space around the CV, such as the disk and/or secondary star wind and nova and DN outbursts, there appears to be sound justification to expect that dust could form in such an environment.

The bremsstrahlung mechanism can operate in astrophysical environments with conditions similar to those found in regions of a CV (i.e., ionized plasmas). It was 
previously excluded as the origin of the IR excess in V592~Cas for a number of reasons discussed in \citet{hoard09} and references therein (see Section~4.1 in that paper).  
Notably, the mass loss rate in the accretion disk wind required to explain the observed level of IR excess via bremsstrahlung would be $\sim5\times10^{-9}$~M$_{\odot}$~yr$^{-1}$, more than 30\% of the inferred steady state mass transfer rate in V592~Cas.  By comparison, 
\citet{proga02}, \citet{noebauer10}, and \citet{puebla11} 
modelled wind mass loss rates for several high mass transfer rate NLs in common with the targets of our survey:\ 
$\lesssim10^{-10}$~M$_{\odot}$~yr$^{-1}$ for RW~Tri, V347~Pup, and IX~Vel, 
and a higher value of 
$\sim1\times10^{-9}$~M$_{\odot}$~yr$^{-1}$ for UX~UMa.
In all four cases, the wind mass loss rate is $\lesssim8$\% of the mass transfer rate.
In fairness, we note that it is currently unclear whether this situation reflects a true generality in the relative values of the mass transfer and disk wind loss rates, or a shortcoming of the disk wind models in reproducing an extremely complex and poorly constrained process.

Nonetheless, we can attempt to ascertain the likely spectral profile shape ($f_{\nu}\propto\nu^{\alpha}\propto\lambda^{-\alpha}$) of a bremsstrahlung component originating in a CV wind by comparing theoretical predictions to the observed IR SED of V592 Cas.
Using the re-extracted photometry for V592 Cas presented here, the observed value of the flux density ratio 
for the IR excess component (i.e., exclusive of the WD, secondary star, and accretion disk)
between the MIPS-24 (23.675~$\mu$m) and IRSPUI-blue (15.8~$\mu$m) bands 
is $f_{\nu,24}/f_{\nu,16}=0.65$, corresponding to a spectral index of $\alpha\approx1.06$\footnote{In \citet{hoard09}, the observed value for this ratio was reported as 0.91, corresponding to $\alpha\approx0.23$; however, this utilized the older flux calibration of the MIPS and IRSPUI-blue data.}.
The bremsstrahlung spectral profile is strongly influenced by the characteristics of the emitting plasma and can exhibit a wide range of frequency dependence in different plasma geometries.
In the simplest case of an optically thin plasma ``slab'' at constant density, the wavelength dependence of bremsstrahlung varies in proportion to the Gaunt factor and is approximately flat ($\alpha\approx0$).  
In a spherically symmetric wind outflow at constant velocity (i.e., with density related to radial distance by $\rho \propto r^{-2}$), the expected bremsstrahlung spectral index in the IR--radio is $\alpha$=0.67--0.6, respectively \citep{wb75}.  
\citet{bc77} showed that bremsstrahlung originating in an accelerating wind (e.g., near the base of the wind in O supergiants) should have a spectral index $\alpha > 0.67$ in the IR, corresponding to a density distribution in the wind of $\rho \propto r^{-\beta}$, where $\beta > 2$ ($\alpha$ and $\beta$ are related by $\alpha=[4\beta - 6]/[2\beta -1]$; also see \citealt{wb75}).
At high optical depth (e.g., in the region where a stellar wind emerges from the photosphere -- see \citealt{bc77} -- or possibly in the innermost regions of a CV accretion disk -- as suggested by \citealt{harrison13}), bremsstrahlung has a blackbody spectral shape (i.e., $\alpha=2$ on the Rayleigh-Jeans tail in the IR; essentially, it is the observed thermal continuum of the accretion disk).

If the observed excess in the IR SED of V592 Cas is due to bremsstrahlung, then it would be most consistent with the case of an accelerating wind with a density distribution of $\rho \propto r^{-2.6}$.  
In the case of the other CVs modelled here (see Section~\ref{s:models}), the flux density ratios of their less prominent IR excess components is $f_{\nu,24}/f_{\nu,16}\approx0.70$--0.75, which yields $\alpha\approx0.9$--0.7, respectively.
This value of $\alpha$ is slightly larger than the expectation for bremsstrahlung from a constant velocity wind, and is consistent with the accelerating wind case, at a lower wind density than for V592 Cas.
However, this is far from a complete picture of this scenario.  The actual bremsstrahlung profile is a function of many additional parameters, including mass flux and terminal velocity of the wind, mean atomic weight of the gas, the Gaunt factor, and so on (see \citealt{wb75} and \citealt{bc77}), and can presumably be a combination of emission from several regions with different conditions.
Ultimately, a component that can, in essence, be shaped to fit any observed data is not helpful in discriminating the actual explanation for the observations (but neither does this shortcoming alone rule out bremsstrahlung as a viable mechanism contributing to the IR SED of NLs).

If the IR emission component is due to bremsstrahlung, then we would also expect its properties (especially overall brightness) to be causally tied to the instantaneous accretion disk properties at any given time, primarily the disk brightness (which is, in turn, directly linked to the rate of mass flow through the disk).  This is true whether the bremsstrahlung originates in a wind from the disk or in the dense regions of the inner disk itself (as proposed by \citealt{harrison13}).  IR emission from circumbinary dust, on the other hand, should be largely independent of stochastic fluctuations in the brightness of the accretion disk around a long-term mean.  In that regard, the behavior of IX~Vel suggests that the IR excess component is more consistent with emission from circumbinary dust:\ even when IR spectra observed at different epochs indicated a significant ($\sim50$\%) change in the mass transfer rate through the accretion disk, the additional IR emission component required to reproduce the corresponding SEDs of IX~Vel remained unchanged (see Section~\ref{s:ixvel_model}).
This appears to be a significant point in favor of the dust scenario.

\subsubsection{The Missing Emission Feature}
\label{s:missing}

Our five targets with {\it Spitzer} IRS spectra that extend past 10~$\mu$m show no evidence for the broad silicate emission feature that is the hallmark of circumstellar dust around isolated WDs (e.g., \citealt{jura09}).  The additional five targets that do not have an IRS spectrum but do have a {\it WISE} W3 (11~$\mu$m) photometric measurement also do not show evidence for enhanced emission in this wavelength region.  We have shown in Section~\ref{s:ir_excess} that the elevated W3 value for V592~Cas in the {\em WISE} All Sky Catalog is, in fact, due to contamination from a nearby star.  Also as noted above, V442~Oph and WX~Ari lack data at $\lambda>8$~$\mu$m, so the presence of silicate emission in these two systems that each have a significant IR excess currently cannot be investigated.

This situation might be inductively construed as lending additional credence to the bremsstrahlung option for explaining the IR excess, since no silicate emission feature would then be expected.
On the other hand, \citet{hoard09,hhs10,hls12} discuss possible scenarios in which the dust grain size distribution (if skewed toward large grains) could weaken the $10$~$\mu$m silicate emission feature so that it is undetectable.  
As a rule of thumb, a spectral feature from
dust at a particular wavelength is only present
if the light-scattering grains are smaller than
that wavelength (e.g., \citealt{spitzer78,khm80,dch06,vh08}).
Hence, while this property has been used to infer that dust grains in WD disks
must have a typical size smaller than
a few microns, it could also suggest that the dust grains in CVs are, on average, larger than $\sim10$~$\mu$m.

It is worth noting that the basic premise, that dust around CVs and single WDs should have similar observational properties, could be flawed, since these two types of object offer different environments and formation scenarios for dust.  
The origin of dust around WDs is established as a ``top-down'' process of collisional grinding of material from a tidally disrupted asteroid that was perturbed inside the Roche limit of the WD \citep{ds02,jura03}.
Yet, despite the fact that the total inferred dust mass in both WDs and CVs is similar (equivalent to a medium--large Solar System asteroid, $\sim10^{21}$--$10^{23}$~g; this work, \citealt{hoard09}, Section~5.6.4 in \citealt{farihi11}), tidal disruption of an asteroid by the WD cannot account for the formation of dust in circumbinary space around a CV.
The scenario for dust formation in CVs would proceed through a ``bottom-up'' process of coagulation or clumping of smaller particles (gas and small dust grains; \citealt{lvw10}) that have escaped the inner binary.
The specific formation path for very large ($\sim10$~$\mu$m) dust grains in circumbinary space around a CV remains an open question.
See additional discussion of this topic in \citet{hoard12}.

A rather fundamental consideration that has been overlooked so far in this discussion is that we would also not expect to observe a strong (or any) 10~$\mu$m silicate emission feature if the chemical composition of the dust is not predominantly silicate. 
For WDs, the dust has been shown to be chemically similar to the composition of the crust and mantle of terrestrial planets \citep{zkm07,fbg11,zkd11,gkf12} and, hence, is silicate-rich.  This is consistent with its origin from a tidally disrupted asteroid. 
In CVs, however, there is no such constraint on the origin of dust.
For example, \citet{ae87} suggest that under some circumstances the WDs and/or accretion disks in CVs might offer natural formation sites for carbon dust grains which could be subsequently expelled by radiation pressure.  However, this process would seem to require the unlikely combination of a cool ($T\lesssim12,000$~K) WD and a bright (hot) accretion disk (also see \citealt{jfk93}).
Nonetheless, observations of the ejecta of Nova Herculis 1991 are consistent with the formation of carbon dust grains \citep{hs94}, while a mix of silicate and carbon/graphite grains are present in six novae observed by \citet{sar95}.  
Counterbalancing this is the possibility that circumbinary dust production in CVs might be similar to that in the upper atmospheres and winds of asymptotic giant branch stars, which produce several times more silicate (mainly SiC) grains than carbon (mainly amorphous graphite) grains (as deduced from compositional analyses of presolar grains in meteorites; \citealt{hz00,draine03}).

In the end, the cases that can currently be made in favor of (or in opposition to) either the bremsstrahlung or dust mechanisms as the origin of the observed IR excesses in CVs are rather ambiguous, largely circumstantial, and in need of stricter observational constraints.

\section{Conclusions}
\label{s:conc}

The presence of an IR excess over the standard model accretion disk at wavelengths longer than $\approx3$--$5$~$\mu$m is ubiquitous in our sample of 12 high mass transfer rate NLs.  
V592~Cas remains the NL with the most significant known IR excess out to wavelengths longer than 20~$\mu$m.
Both V442~Oph and WX~Ari have IR excesses brighter than even that of V592~Cas out to 7.9~$\mu$m (IRAC-4), but we currently lack longer wavelength data to further constrain and characterize the source of the excess in these two systems.
The SED of the excess component is similar in shape in all cases, and can be modelled with a circumbinary dust disk.  We note that, in the context of modelling the IR SEDs of CVs, using a blackbody to represent the secondary star can lead to false-negatives for detection of IR excess, since this practice overestimates the brightness of a late-type star at long wavelengths.  Despite the fact that it introduces complexity to the modeling process, it is advisable to use models that are as accurate and realistic as possible to represent all of the SED components.

In our sample, there is no significant correlation between the level of IR excess and the orbital period.
Thus, while the presence of an IR excess is likely linked to the overall high mass transfer rate shared by all of our targets ($\gtrsim10^{-9}$~M$_{\odot}$~yr$^{-1}$), its relative strength from system to system does not appear to be strongly dependent on the specific value of mass transfer rate over the ranges of orbital period (0.11--0.45~d; see Table~\ref{t:targets}) and mass transfer rate ($\sim1$--$100\times10^{-9}$~M$_{\odot}$~yr$^{-1}$; see Tables~\ref{t:models1} and \ref{t:models2}) spanned by our sample of NLs.
Coupled with the fact that the modelled total dust masses needed to reproduce the observed IR excesses are significantly smaller than the predicted threshold necessary to influence the secular evolution of CVs (also see discussion below), 
this implies that the presence of dust is not a dominant factor in driving that evolution.
This conclusion is only valid if the mass transfer rate at a given orbital period is primarily determined by secular evolution (and, of course, if the mass transfer rates returned by the various accretion disk models are truly representative of the situation in these CVs).

On the other hand, the IR excess and system inclination show a statistically significant correlation.
The increase in excess at 7.9~$\mu$m with increasing inclination amounts to an $\approx20$\% change from an inclination of $30^{\circ}$ (near face-on) to $90^{\circ}$ (edge-on).
It is tempting to ascribe this
relatively shallow trend to the changing view of the accretion disk at different inclinations (i.e., the hot, inner disk is increasingly self-occulted as inclination increases, leaving the accretion disk contribution to the SED to be dominated by the cooler outer disk and disk rim).
However, comparison of model accretion disk SEDs computed at different inclinations suggests that this effect is likely not the sole factor in producing the observed level of change in IR excess.

There is an observational test that might be able to discriminate the location of the component producing the IR excess. The IR emission from circumbinary dust should not be affected by the primary eclipse in high inclination CVs, whereas IR emission originating from the inner binary (e.g., via bremsstrahlung or extremely cool material associated with the outer accretion disk) would likely be eclipsed to some extent, barring an extreme configuration in which the IR emission originates at large vertical distances above the accretion disk.  While the latter scenario could be envisioned for free-free emission in an outflowing disk wind, the analysis presented here (see Section~\ref{s:brems}) implies that if the IR excess originates in a wind, then it most likely comes from the dense, accelerating base of the wind.  Time-resolved mid-IR ($\lambda\gtrsim5$~$\mu$m) observations of the eclipsing NLs in our sample (RW~Tri, V347~Pup) or other high inclination NLs confirmed to have an IR excess would be helpful in this regard.
Overall, however, the presence of a strong IR excess appears to go hand-in-hand with a high mass transfer rate, possibly via the formation of circumbinary dust in material that escapes the inner binary, as discussed in \citet{hoard09}.  
This is not to say that CVs with lower mass transfer rates would be necessarily devoid of dust, but if present, then it is likely in a smaller amount, producing a weaker IR excess that is difficult to detect.

None of the CVs we observed show evidence in their SEDs for the presence of the 10~$\mu$m emission feature that is the hallmark of the silicate dust found around numerous isolated WDs.  There are a number of possible reasons why this feature could be absent even when dust is present in a CV (see Section~\ref{s:missing}), including predominantly large grain size and differences in chemical composition of the dust.
The dust grain size has a strong effect on the total mass of dust required to produce an IR excess of a given amplitude.
Thus, the lack of silicate emission at 10~$\mu$m has potential implications for the amount of dust that could be present around these NLs (which, in turn, has implications for the potential effect of dust on the secular evolution of CVs).
The total mass of silicate dust in our models that is required to reproduce the observed IR excesses in V592~Cas, IX~Vel, UX~UMa, and RW~Sex is $\approx$1--2$\times10^{21}$~g assuming a characteristic dust grain radius of $r_{\rm grain}=1$~$\mu$m (see Table~\ref{t:models1}).  
For the case of large grains ($r_{\rm grain}\gtrsim5$~$\mu$m) corresponding to the absence of the 10~$\mu$m silicate emission feature, the total dust masses would be larger by a factor of $\gtrsim5$.
At the low end ($r_{\rm grain}\approx5$~$\mu$m), the total mass is still insignificant compared to the prediction of $\gtrsim10^{28}$~g required to influence CV secular evolution \citep{taam03,willems07}.
Even at a {\it reductio ad absurdum} limit of centimeter-scale dust ``pebbles'' ($r_{\rm grain}\approx10^{4}$~$\mu$m), the total dust mass is still 3 orders of magnitude too small to be important in this regard.
While the absence of the 10~$\mu$m silicate emission feature does not necessarily imply the absence of dust, it nonetheless remains true -- and should be considered an observational priority -- that detection of this feature in any quiescent system would be compelling evidence for the presence of dust in CVs.

\acknowledgements{
This work is based on observations made with the 
{\it Spitzer Space Telescope}, which is operated by the Jet 
Propulsion Laboratory (JPL), California Institute of Technology (Caltech), 
under a contract with the National Aeronautics and Space 
Administration (NASA).
Support for this work was provided by NASA.
We acknowledge with thanks the variable star observations from the AAVSO International Database contributed by observers worldwide and used in this research.
This work is also based on data, data products, and other resources obtained from: 
(i) The Two Micron All Sky Survey (2MASS), a joint project of the University of Massachusetts and the Infrared Processing and Analysis Center (IPAC)/Caltech, funded by NASA and the National Science Foundation (NSF).
(ii) NASA's Astrophysics Data System.  
(iii) The NASA/IPAC Infrared Science Archive (IRSA), which is operated by JPL/Caltech, under a contract with NASA.
(iv) The SIMBAD database, operated at CDS, Strasbourg, France.
(v) The {\it Wide-field Infrared Survey Explorer} ({\it WISE\/}), which is a joint project of the University of California, Los Angeles, and JPL/Caltech, funded by NASA. 
}

{\it Facilities:} 
\facility{AAVSO}, 
\facility{Spitzer}, 
\facility{WISE}

\newpage

\include{table_targets}

\newpage
\include{table_log}

\newpage
\include{table_phot}

\newpage

\include{table_corrs}

\newpage
\include{table_models_part1}

\newpage
\include{table_models_part2}

\newpage

\newpage

\appendix

\section{Concise Histories of Targets}

This appendix presents concise summaries of observational characteristics for the targets of our study, compiled from the literature.  While not strictly necessary for interpreting our results, this information provides context for it.

\subsection{V592 Cas}
\label{s:app_v592cas}

V592~Cas was discovered by \citet{greenstein70}. Its blue color and spectral characteristics identified it as a NL. Its optical spectrum displays strong emission from \ion{He}{2}~4686~\AA\ and the \ion{C}{3}/\ion{N}{3}/\ion{O}{2} blend at $\approx4640$--$4650$~\AA\footnote{We note in passing that although this broad emission feature found in many NL spectra is commonly referred to as the ``Bowen blend'', only the \ion{N}{3} emission components actually result from the Bowen flourescence cascade \citep{mct75,sfk89}.}, along with weak Balmer and \ion{He}{1} emission lines \citep{downes95}. \citet{huber98} performed the first time-resolved optical spectroscopic study of V592~Cas and used radial velocity measurements to determine an initial estimate of the orbital period, $P_{\rm orb} = 2.47$~hr. This was later refined to $P_{\rm orb} = 2.76$~hr by \citet{taylor98} and \citet{witherick03}. Although photometric observations show that it does not eclipse (inclination of $i=28^{\circ}$; \citealt{huber98}), V592~Cas does exhibit both positive and negative permanent superhumps \citep{taylor98}.  These are believed to be caused by precession of an eccentric accretion disk and retrograde precession of a tilted accretion disk, respectively (e.g., \citealt{skillman98,kraicheva99,kim09}).  Optical and UV spectroscopic observations demonstrated the presence of a wind, which manifests primarily as a P~Cygni profile in the emission lines \citep{witherick03,prinja04,kafka09}. \citet{kafka09} showed that the wind is variable and reaches velocities up to $v_{\rm wind}\sim5000$~km~s$^{-1}$ (which is greater than the WD escape velocity).

\subsection{IX Vel}

IX~Vel was serendipitously discovered during a survey of OB stars \citep{garrison77} and subsequently classified as a NL \citep{garrison84}.  It displays low amplitude variability of $\sim0.1$--0.5~mag on timescales ranging from minutes to years, and very broad, shallow Balmer absorption lines with emission cores.  Photometric observations by \citet{wargau84} revealed flickering between $m_{\rm pg}=9.1$ and 10.0~mag, but their IR photometry showed no obvious eclipse, setting an upper limit to the inclination of roughly 65--70$^{\circ}$.  
\citet{linnell07} fit a BINSYN model to the far-UV (900--1700~\AA) SED of IX~Vel (see Table~\ref{t:models1}).  They found that the radial temperature profile of the accretion disk follows the standard model outside the innermost disk ($r_{\rm disk} > 4$~$R_{\rm wd}$).  The preliminary UV spectroscopic search for circumbinary disks by \citet{belle04} found no indication of absorption features related to circumbinary material around IX~Vel.

\subsection{QU Car}

Although generally classified with the NLs, QU~Car is also noted as a ``V~Sge'' star.  These objects have a massive WD and an evolved secondary star that allows very high mass transfer rates and leads to sustained nuclear burning on the surface of the WD \citep{steiner98}.  QU~Car itself displays UV and optical nebular emission lines that likely originate in circumbinary material fed by the high velocity wind from the inner binary, and has been proposed as a prime candidate to become a Type Ia supernova \citep{kafka08,kafka12}.  Although its distance is somewhat uncertain, even at lower limit estimates QU~Car is likely the intrinsically most luminous NL \citep{drew03}.  \citet{linnell08b} fit a BINSYN model to the far-UV--optical (900--3000~\AA) SED of QU~Car (see Table~\ref{t:models1}), and found that the hottest inner region of the disk is not well-represented by a standard model accretion disk.  The preliminary UV spectroscopic search for circumbinary disks by \citet{belle04} found no indication of absorption features related to circumbinary material around QU~Car.

\subsection{RW Sex}
\label{s:app_rwsex}

\citet{greenstein82} performed the first in-depth, multi-wavelength investigation of RW~Sex, noting weak X-ray emission, P~Cygni profiles in the UV lines with terminal velocities on the order of WD escape velocity, and near-UV variability on timescales of several minutes.  They suggest a substantial wind from the accretion disk is present (which could provide a source of raw material for the formation of dust).  
The blue optical spectrum of RW~Sex shows broad, deep Balmer absorption lines with emission cores, and weak \ion{He}{2} and \ion{C}{3}/\ion{N}{3} emission lines \citep{bolick87}. 
Optical radial velocity curves constructed by \citet{beuermann92a} identified these emission cores as originating on the heated inner face of the secondary star.  \citet{linnell10} fit a BINSYN model to the far-UV--optical (900--3000~\AA) SED of RW~Sex (see Table~\ref{t:models1}), and found that the hottest inner region of the disk is characterized by a temperature gradient shallower than the standard model (i.e., the inner disk remains hotter with increasing distance than the standard model).  \citet{greenstein82} also found the SED of RW~Sex to be flattened in the UV ($\lambda<5000$~\AA).

\subsection{TT Ari}

TT~Ari was found by \citet{smak69} to show quasiperiodic and periodic variability on timescales of minutes to hours, as well as rapid flickering.
\citet{cowley75} subsequently confirmed it as a CV through time-resolved spectroscopic observations.  TT~Ari is notable for displaying a wide range of variability on short timescales (e.g., \citealt{smak69,mard80}), bright and faint state transitions (i.e., VY~Scl behavior -- see Section~\ref{s:app_vyscl}; e.g., \citealt{boris99,kraicheva99,melikian10}), and both positive and negative superhumps (at different times).  Emission line P~Cygni profiles have been observed in its UV and optical spectra, likely indicative of an outflowing accretion disk wind \citep{melikian04,hutchings07}.

\subsection{RW Tri}

RW~Tri is one of the earliest known CVs, discovered as an eclipsing binary more than 75 years ago \citep{protitch37,protitch38}.  \citet{groot04} modelled RW~Tri using optical (3600--7000~\AA) spectrophotometry, and found a radial temperature profile consistent with a standard steady state accretion disk \citep{FKR02}.  \citet{rutten92} reached the same conclusion via broad-band eclipse mapping.  In a search for circumbinary material around several CVs, \citet{dubus04} found no need for an additional component to explain the near-IR SED of RW~Tri.

\subsection{V347 Pup}

\citet{buckley90} used photometric and spectroscopic observations to identify V347~Pup as a deeply ($\sim3$~mag), but only partially (V-shaped profile), eclipsing NL.  They noted that it is also a transient hard X-ray source, the counterpart to the {\it HEAO~1} source 4U~0608$-$49.  The accretion disk in V347~Pup appears to be at least partially self-occulting (e.g., see \citealt{mrb94}), suggesting that the contribution of the WD and hot inner disk to the system spectrum would likely be weaker than would otherwise be expected.  \citet{thoroughgood05} made estimates of the stellar component masses and radii in V347~Pup, based on a time-resolved spectroscopic and photometric investigation.  They reached a tentative conclusion that the secondary star is evolved.

\subsection{UX UMa}
\label{s:app_uxuma}

UX~UMa is an archetype of the NLs, and has been extensively studied since the mid-20th century (e.g., \citealt{walker53,johnson54,walker54}).  \citet{warner72} and \citet{nather74} found low amplitude, $\approx29$~s oscillations in the light curve of UX~UMa, which they attributed to pulsations of the WD. \citet{knigge98} recovered these oscillations in {\it Hubble Space Telescope} time-series UV spectrophotometry, and discuss their origin in terms of a compact region of the inner accretion disk.  \citet{linnell08a} fit a BINSYN model to the far-UV--optical (900--8000~\AA) SED of UX~UMa (see Table~\ref{t:models1}).  Although they found it challenging to produce a model consistent with all available observations, they note similarities of the accretion disk component with those in their models for IX~Vel and QU~Car; namely, that the radial temperature profile of the inner disk must be shallower (i.e., cooler) than the standard model (a feature also noted by \citealt{noebauer10} in their Monte Carlo radiative transfer models of UX~UMa).  Nonetheless, \citet{linnell08a} conclude that a standard accretion disk model provides a reasonable fit to the observed SED of UX~UMa. However, the model residuals from data sets obtained at different times suggest that the mass transfer rate in UX~UMa is somewhat variable.
\citet{neustroev11} also observed differences between optical spectra of UX~UMa obtained in 1999 and 2008 that could be linked to a change in overall mass transfer rate at the two epochs.

\subsection{VY Scl}
\label{s:app_vyscl}

VY~Scl was first discovered in 1960 as an irregularly variable, faint blue star \citep{haro60}.  Its long-term photometric behavior (long stretches of stable brightness near 13th magnitude interspersed with drops to 18th magnitude lasting about 200 days) was noted as unique at the time \citep{warner74}.  While initially classified as an R~Coronae~Borealis star \citep{pismis72}, \citet{kraft64} suggested that VY~Scl might be a Z~Camelopardalis type DN\footnote{Following normal outbursts, the Z~Cam subclass of DN spends months to years in an intermediate brightness ``standstill'' state about 1~mag fainter than the outburst peak before dropping back to the quiescent state \citep{warner03}.} that is mostly stuck in the ``standstill'' state, only dropping occasionally to its normal, faint quiescent state.  Almost a decade later, \citet{burrell73} refuted the R~CrB classification for VY~Scl, and supported the Z~Cam DN classification.  However, they noted that the ``chief difficulty'' of this classification is that ``VY Scl shows irregular minima, whereas [dwarf novae] undergo outbursts''.  Hence, even with no comparable systems known at the time, \citet{burrell73} (as well as \citealt{kraft64} and \citealt{warner74}) still identified in VY~Scl the characteristic long-term photometric behavior that would later become the hallmark of the VY~Scl stars.
Members of this subtype of NL 
display intermittent, unpredictable drops to a faint state up to several magnitudes below their normal bright state level that last from weeks to months to years \citep{leach99,HK04}.  This is believed to be caused by throttling of the accretion flow, likely due to solar-like cyclic magnetic activity on the secondary star \citep{livio94}, or possibly a feedback loop in which irradiation of the secondary star by the hot inner accretion disk increases the mass transfer rate, which, in turn, inflates the outer disk and thereby reduces irradiation of the secondary star \citep{wu95}.  Despite its status as a CV archetype, VY~Scl itself is relatively little-studied, with only a few in-depth observational investigations in the past several decades \citep{hutchings84,pais00,hamilton08}.

\subsection{V3885 Sgr}

V3885~Sgr was discovered as an emission line object \citep{bidelman68}; shortly thereafter, \citet{cowley69} published its complex optical spectrum, noting emission lines of \ion{H}{1}, \ion{He}{1}, and multiply ionized C, O, and Si.  \citet{ribeiro07} localized the main source of flickering in the H$\alpha$ emission line to the irradiated face of the secondary star rather than the accretion disk.
V3885~Sgr provided the first unambiguous detection of spiral waves in the accretion disk of a NL \citep{hartley05}; frequently found in DNe during outburst, the spiral waves originate from gravitational compression of the outer regions of an elliptical accretion disk by the WD as the disk rotates, which propagate inward as spiral shocks (e.g., see discussion in \citealt{hartley05}).  

V3885~Sgr also has the distinction of being the first and, to date, only known NL with a reproducible radio detection \citep{kording11}; as such, it joins a small number of CVs with confirmed radio emission (primarily magnetic systems and outbursting classical and recurrent novae, but 
also including the DN SS~Cyg during outburst; \citealt{krk08}).  Radio emission possibly indicates the presence of jets producing optically thin synchrotron radiation.  \citet{linnell09} fit a BINSYN model to the far-UV (900--1700~\AA) SED of V3885~Sgr (see Table~\ref{t:models2}), and found good agreement with a standard model accretion disk.  The preliminary UV spectroscopic search for circumbinary disks by \citet{belle04} found no indication of absorption features related to circumbinary material around V3885~Sgr.

\subsection{V442 Oph}

V442~Oph was identified as a NL by \citet{szkody80} based on its photometric and spectroscopic properties. They noted that the \ion{He}{2} $\lambda4686$ emission is comparable in strength to \ion{H}{1}. A follow-up investigation by \citet{shafter83} utilized photometry and spectroscopy in the IR, optical, and UV. They measured an orbital period of $P_{\rm orb}=0.1403$~d (=7.13 cycles~d$^{-1}$) from \ion{H}{1} and \ion{He}{2} $\lambda4686$ radial velocity curves and found a roughly sinusoidal variation in the J-band light curve consistent with this period (presumably caused by the changing visibility of the irradiation-heated inner face of the secondary star). However, extensive time series optical photometry did not reveal any evidence of an eclipse, implying a system inclination of $i\lesssim65^{\circ}$.  \citet{hoard00} suggested that V442~Oph shared many of the observational characteristics of the SW~Sex stars.  They used extensive H$\alpha$ radial velocity measurements to demonstrate that the true orbital period of V442~Oph is 0.12435(7)~d (=8.06 cycles~d$^{-1}$), with the longer \citet{shafter83} period being a 1-day alias.  \citet{diaz01} subsequently refined the orbital period to 0.1243301(9)~d using additional H$\alpha$ radial velocity measurements.  \citet{patterson02} identified low amplitude, persistent superhumps (both positive and negative) in V442~Oph, and suggested that persistent superhumps might be present in all SW~Sex stars.

\subsection{WX Ari}

There are few dedicated observational investigations of WX~Ari.  \citet{beuermann92b} obtained time-resolved optical spectra and determined its orbital period from radial velocity measurements.  They suggested that WX~Ari was a member of the then recently proposed SW~Sex class, but noted that it did not appear to eclipse (all known SW~Sex stars at the time were deeply eclipsing).  In a short note, \citet{hellier94} concluded from several hours of V-band photometry that WX~Ari did not eclipse.  However, \citet{gil00} obtained a longer set of R-band photometry and found that WX~Ari is at an intermediate inclination ($i\sim72^{\circ}$) with shallow ($\Delta R\sim0.15$~mag) partial eclipses. \citet{ballouz09} favor a lower inclination ($i\sim60^{\circ}$) based on model fits to the UV spectrum of WX~Ari.

\end{document}

%% file: table_targets.tex
\begin{deluxetable}{llll}
\tablewidth{0pt}
\tabletypesize{\scriptsize}
\tablecaption{Target Information \label{t:targets}} 
\tablehead{
\colhead{Target} & 
\colhead{Distance} &
\colhead{Orbital Period} &
\colhead{Inclination} \\
\colhead{ } & 
\colhead{($d$; pc)} &
\colhead{($P_{\rm orb}$; d)} &
\colhead{($i$; $^\circ$)} 
}
\startdata
V592 Cas  & 380 [1]                         & 0.115063(1) [2]     & $28^{+11}_{-10}$ [2] \\ [2pt]
IX Vel    &  96(9) [3]                      & 0.193929(2) [4]     & 57(2) [3] \\ [2pt]
QU Car    & $\sim610$ [5]                   & 0.454(14) [6]       & 40--55 [5] \\ [2pt]
RW Sex    & 150, 289 [7]                    & 0.24507(20) [8]     & 34(6) [7] \\ [2pt]
TT Ari    & 335(50)[9]                      & 0.13755040(17) [10] & 29(6) [10] \\ [2pt]
RW Tri    & 330(40) [11], 341(+38,-31) [12] & 0.231883297 [11]    & 75(+5,-8) [13] \\ [2pt]  
V347 Pup  & 470(130) [14]                   & 0.231936060(6) [14] & 84(2) [14] \\ [2pt]
UX UMa    & 250--345 [15], 345(34) [13]     & 0.196671278(2) [16] & 70.2(2) [15], 71.0(6) [13], 65--75 [16] \\ [2pt]
V3885 Sgr & 110(30) [17]                    & 0.20716071(22) [18] & 65(2) [17] \\ [2pt]
VY Scl    & 536--620 [19]                   & 0.232(3) [20]       & 30(10) [20], 41 [19] \\ [2pt]
V442 Oph  & 183 [21]                        & 0.12435(7) [22]     & 70(5) [21,22] \\  [2pt]
WX Ari    & 468 [21]                        & 0.13935119(3) [23]  & $\approx60$ [21], $\approx72$ [23] 
\enddata
\tablerefs{~Listed in brackets with each parameter value: 
 1 = this work; 
 2 = \citet{hoard09};
 3 = \citet{linnell07};
 4 = \citet{beuermann90};
 5 = \citet{linnell08b};
 6 = \citet{gilliland82};
 7 = \citet{linnell10} -- the smaller distance is from Bailey's method and yields the favored model, while the larger distance is from the {\it Hipparcos} parallax and has a $1\sigma$ lower limit of 169~pc;
 8 = \citet{hillwig04};
 9 = \citet{boris99};
10 = \citet{wu02};
11 = \citet{groot04};
12 = \citet{mcarthur99} -- distance from {\it Hubble Space Telescope} Fine Guidance Sensor trigonometric parallax;
13 = \citet{noebauer10};
14 = \citet{thoroughgood05};
15 = \citet{linnell08a};
16 = \citet{neustroev11};
17 = \citet{linnell09} -- distance is from the {\it Hipparcos} parallax;
18 = \citet{ribeiro07};
19 = \citet{hamilton08};
20 = \citet{pais00};
21 = \citet{ballouz09};
22 = \citet{hoard00};
23 = \citet{gil00}.
}
\end{deluxetable}

%% file: table_log.tex
\begin{deluxetable}{lllllllll}
\tablewidth{0pt}
\tabletypesize{\scriptsize}
\rotate
\tablecaption{Observation Log \label{t:log}} 
\tablehead{
\colhead{ } & 
\colhead{Instrument/} & 
\colhead{ } & 
\multicolumn{3}{l}{Observation Date:} &
\colhead{Processing} &
\colhead{{\it Spitzer}} &
\colhead{{\it Spitzer}} \\
\colhead{Target} & 
\colhead{Survey} & 
\colhead{Band(s)} & 
\colhead{UT} &
\colhead{Year} &
\colhead{JD-2450000} &
\colhead{Version} &
\colhead{Program} &
\colhead{AOR} 
}
\startdata
V592 Cas  & 2MASS          & JHKs  & 2000 Sep 18           & 2000.742           & 1815.92       & All Sky  & \nodata & \nodata \\
          & IRAC   & 1--4  & 2005 Aug 18           & 2005.632           & 3601.25       & S18.25.0 & 20221   & 14408960 \\
          & IRSPUI & blue  & 2006 Jan 16           & 2006.046           & 3752.17       & S18.18.0 & 20221   & 14412032 \\
          & MIPS   & 24    & 2006 Feb 21           & 2006.143           & 3787.88       & S18.12.0 & 20221   & 14412288 \\
          & WISE           & 1--4  & 2010 Jan 07 -- Aug 06 & 2010.019--2010.597 & 5204--5415 & All Sky  & \nodata & \nodata \\ [5.0pt]

IX Vel    & 2MASS          & JHKs       & 1999 Dec 20           & 1999.969           & 1532.76       & All Sky  & \nodata & \nodata \\
          & IRAC   & 1--4       & 2005 Jun 10           & 2005.442           & 3532.03       & S18.25.0 & 20221   & 14412800 \\
          & MIPS   & 24         & 2005 Dec 07           & 2005.936           & 3712.32       & S18.12.0 & 20221   & 14413568 \\
          & IRSPUI & blue       & 2005 Dec 18           & 2005.966           & 3723.43       & S18.18.0 & 20221   & 14413312 \\
          & IRS    & SL1\&2     & 2006 Feb 02           & 2006.091           & 3768.78       & S18.18.0 & 20221   & 14413056 \\
          & IRS    & SL1\&2,LL2 & 2009 Apr 06           & 2009.265           & 4928.35       & S18.18.0 & 50068   & 25368832 \\
          & IRS\tablenotemark{a} & SH & 2009 Apr 06     & 2009.265           & 4928.35       & S18.18.0 & 50068   & 25368832 \\
          & WISE           & 1--4       & 2010 Jan 07 -- Aug 06 & 2010.019--2010.597 & 5204--5415 & All Sky  & \nodata & \nodata \\ [5.0pt]

QU Car    & 2MASS          & JHKs       & 2000 Jan 24           & 2000.067           & 1567.84       & All Sky  & \nodata & \nodata \\
          & IRAC   & 1--4       & 2009 Mar 18           & 2009.212           & 4908.78       & S18.25.0 & 50068   & 25369856 \\
          & IRS    & SL1\&2     & 2009 Apr 11           & 2009.277           & 4932.51       & S18.18.0 & 50068   & 25369344 \\
          & IRSPUI & red        & 2009 Apr 11           & 2009.277           & 4932.53       & S18.18.0 & 50068   & 25369600 \\
          & WISE           & 1--4       & 2010 Jan 07 -- Aug 06 & 2010.019--2010.597 & 5204--5415 & All Sky  & \nodata & \nodata \\ [5.0pt]

RW Sex    & 2MASS          & JHKs       & 1999 Jan 15           & 1999.042           & 1193.74       & All Sky  & \nodata & \nodata \\
          & IRS    & SL1\&2,LL2 & 2009 Jan 25           & 2009.069           & 4856.60       & S18.18.0 & 50068   & 25370112 \\
          & IRAC   & 1--4       & 2009 Feb 04           & 2009.097           & 4866.95       & S18.25.0 & 50068   & 25370368 \\
          & WISE           & 1--4       & 2010 Jan 07 -- Aug 06 & 2010.019--2010.597 & 5204--5415 & All Sky  & \nodata & \nodata \\ [5.0pt]

TT Ari    & 2MASS          & JHKs       & 1998 Sep 24           & 1998.732           & 1080.88       & All Sky  & \nodata & \nodata \\
          & IRS    & SL1\&2     & 2009 Mar 08           & 2009.185           & 4899.25       & S18.18.0 & 50068   & 25371648 \\
          & IRSPUI & red        & 2009 Mar 08           & 2009.186           & 4899.29       & S18.18.0 & 50068   & 25371904 \\
          & IRAC   & 1--4       & 2009 Mar 12           & 2009.196           & 4903.08       & S18.25.0 & 50068   & 25372160 \\
          & WISE           & 1--4       & 2010 Jan 07 -- Aug 06 & 2010.019--2010.597 & 5204--5415 & All Sky  & \nodata & \nodata \\ [5.0pt]

RW Tri    & 2MASS          & JHKs       & 1997 Nov 08           & 1997.855           & \phn760.79       & All Sky  & \nodata & \nodata \\
          & IRS    & SL1\&2     & 2009 Mar 07           & 2009.182           & 4898.04       & S18.18.0 & 50068   & 25370880 \\
          & IRSPUI & red        & 2009 Mar 07           & 2009.182           & 4898.11       & S18.18.0 & 50068   & 25371392 \\
          & IRAC   & 1--4       & 2009 Mar 12           & 2009.196           & 4903.07       & S18.25.0 & 50068   & 25371136 \\
          & WISE           & 1--4       & 2010 Jan 07 -- Aug 06 & 2010.019--2010.597 & 5204--5415 & All Sky  & \nodata & \nodata \\ [5.0pt]

V347 Pup  & 2MASS          & JHKs       & 1999 Nov 10           & 1999.861           & 1492.85       & All Sky  & \nodata & \nodata \\
          & IRS    & SL2        & 2008 Dec 12           & 2008.949           & 4813.13       & S18.18.0 & 50068   & 25373184 \\
          & IRSPUI & blue       & 2008 Dec 12           & 2008.949           & 4813.16       & S18.18.0 & 50068   & 25373440 \\
          & IRAC   & 1--4       & 2008 Dec 19           & 2008.969           & 4820.32       & S18.25.0 & 50068   & 25373696 \\
          & WISE           & 1--4       & 2010 Jan 07 -- Aug 06 & 2010.019--2010.597 & 5204--5415 & All Sky  & \nodata & \nodata \\ [5.0pt]

UX UMa    & 2MASS          & JHKs       & 2000 Mar 19           & 2000.214           & 1622.85       & All Sky  & \nodata & \nodata \\
          & MIPS   & 24         & 2007 Jul 06           & 2007.513           & 4287.76       & S18.12.0 & 40204   & 22676224 \\
          & IRAC   & 1--4       & 2009 Jan 27           & 2009.075           & 4858.78       & S18.25.0 & 50068   & 25372928 \\
          & IRS    & SL2        & 2009 Mar 06           & 2009.179           & 4896.77       & S18.18.0 & 50068   & 25372416 \\
          & IRSPUI & blue       & 2009 Mar 06           & 2009.179           & 4896.81       & S18.18.0 & 50068   & 25372672 \\
          & WISE           & 1--4       & 2010 Jan 07 -- Aug 06 & 2010.019--2010.597 & 5204--5415 & All Sky  & \nodata & \nodata \\ [5.0pt]

VY Scl    & 2MASS          & JHKs       & 1999 Jun 12           & 1999.447           & 1341.90       & All Sky  & \nodata & \nodata \\
          & IRSPUI & blue       & 2008 Dec 12           & 2008.949           & 4813.02       & S18.18.0 & 50068   & 25374976 \\
          & IRS\tablenotemark{a} & SL2 & 2008 Dec 13    & 2008.951           & 4813.99       & S18.18.0 & 50068   & 25374720 \\
          & IRAC   & 1--4       & 2008 Dec 18           & 2008.966           & 4819.31       & S18.25.0 & 50068   & 25375232 \\
          & WISE           & 1--4       & 2010 Jan 07 -- Aug 06 & 2010.019--2010.597 & 5204--5415 & All Sky  & \nodata & \nodata \\ [5.0pt]

V3885 Sgr & 2MASS          & JHKs       & 1999 Sep 15           & 1999.707           & 1436.57       & All Sky  & \nodata & \nodata \\
          & IRAC   & 1--4       & 2008 Oct 30           & 2008.831           & 4769.96       & S18.25.0 & 50068   & 25374208 \\
          & WISE           & 1--4       & 2010 Jan 07 -- Aug 06 & 2010.019--2010.597 & 5204--5415 & All Sky  & \nodata & \nodata \\ [5.0pt]

V442 Oph  & 2MASS          & JHKs       & 1998 Apr 27           & 1998.321           & \phn930.88       & All Sky  & \nodata & \nodata \\
          & IRAC   & 1--4       & 2005 Sep 15           & 2005.708           & 3628.91       & S18.25.0 & 20221   & 14409216 \\
          & WISE           & 1--4       & 2010 Jan 07 -- Aug 06 & 2010.019--2010.597 & 5204--5415 & All Sky  & \nodata & \nodata \\ [5.0pt]

WX Ari    & 2MASS          & JHKs       & 1999 Dec 17           & 1999.961           & 1529.62       & All Sky  & \nodata & \nodata \\
          & IRAC   & 1--4       & 2005 Aug 19           & 2005.634           & 3601.92       & S18.25.0 & 20221   & 14409472 \\
          & WISE           & 1--4       & 2010 Jan 07 -- Aug 06 & 2010.019--2010.597 & 5204--5415 & All Sky  & \nodata & \nodata \\ 
\enddata
\tablenotetext{a}{These data are not used in this work due to low S/N.}
\end{deluxetable}

%% file: table_phot.tex
\begin{deluxetable}{lllllllllllllllll}
\tablewidth{0pt}
\tabletypesize{\scriptsize}
\rotate
\setlength{\tabcolsep}{0.050in} 
\tablecaption{Infrared Photometry\tablenotemark{a} \label{t:phot}} 
\tablehead{
\colhead{ } & 
\colhead{2MASS} & 
\colhead{2MASS} & 
\colhead{2MASS} & 
\colhead{{\it WISE}} &
\colhead{IRAC} &
\colhead{IRAC} &
\colhead{{\it WISE}} &
\colhead{IRAC} &
\colhead{IRAC} &
\colhead{{\it WISE}} &
\colhead{IRSPUI} &
\colhead{{\it WISE}} &
\colhead{IRSPUI} &
\colhead{MIPS} &
\colhead{ } &
\colhead{ } \\
\colhead{Target} & 
\colhead{J} & 
\colhead{H} & 
\colhead{Ks} &
\colhead{W1} &
\colhead{1} &
\colhead{2} &
\colhead{W2} &
\colhead{3} &
\colhead{4} &
\colhead{W3} &
\colhead{blue} &
\colhead{W4} &
\colhead{red} &
\colhead{24} &
\colhead{Methods\tablenotemark{b}} &
\colhead{Notes} \\
\colhead{ } & 
\colhead{1.235} & 
\colhead{1.662} & 
\colhead{2.159} &
\colhead{3.353} &
\colhead{3.550} &
\colhead{4.493} &
\colhead{4.603} &
\colhead{5.731} &
\colhead{7.872} &
\colhead{11.561} &
\colhead{15.8} &
\colhead{22.088} &
\colhead{22.3} &
\colhead{23.68} &
\colhead{ } &
\colhead{ } 
}
\startdata
V592 Cas  &      19.27 &     12.82 &      8.88 &      4.42 &      4.15 &      2.92 &      2.73 &      1.98 &      1.28 &      1.08 &       0.50 &   $<$2.07 & \nodata   &      0.43 & I~I~A & 1,2 \\
          &  $\pm$0.53 & $\pm$0.44 & $\pm$0.25 & $\pm$0.12 & $\pm$0.19 & $\pm$0.14 & $\pm$0.07 & $\pm$0.11 & $\pm$0.06 & $\pm$0.08 &  $\pm$0.04 &           &           & $\pm$0.03 &       & \\
          &[$22.28\pm1.85$]&[$13.95\pm1.32$]&[$9.35\pm0.91$]& &        &           &           &           &           & [$0.74\pm0.02$] &      &           &           &           &       & \\ [6.0pt]

IX Vel    &     359.16 &    261.76 &    196.24 &     97.70 &     90.37 &     62.91 &     58.53 &     42.18 &     26.01 &     13.77 &       7.28 &      4.32 & \nodata   &      3.91 & I~A~A & \nodata \\ 
          & $\pm$12.44 & $\pm$7.95 & $\pm$5.06 & $\pm$2.48 & $\pm$4.08 & $\pm$2.83 & $\pm$1.40 & $\pm$1.91 & $\pm$1.18 & $\pm$0.34 &  $\pm$0.55 & $\pm$1.08 &           & $\pm$0.25 &       & \\ [6.0pt]

QU Car    &      65.12 &     46.89 &     34.77 &     19.05 &     15.30 &     10.72 &     11.30 &      7.52 &      4.84 &      2.80 &  \nodata   &   $<$1.28 &      0.73 & \nodata   & I~I~- & \nodata \\ 
          &  $\pm$1.75 & $\pm$1.29 & $\pm$0.93 & $\pm$0.48 & $\pm$0.77 & $\pm$0.71 & $\pm$0.26 & $\pm$0.40 & $\pm$0.23 & $\pm$0.10 &            &           & $\pm$0.05 &           &       & \\ [6.0pt]

RW Sex    &     118.60 &     89.60 &     62.52 &     32.14 &     30.13 &     20.85 &     18.65 &     13.81 &      8.45 &      4.39 &  \nodata   &   $<$1.94 & \nodata   & \nodata   & I~-~- & \nodata \\ 
          &  $\pm$3.54 & $\pm$2.66 & $\pm$1.70 & $\pm$0.86 & $\pm$1.46 & $\pm$0.99 & $\pm$0.44 & $\pm$0.66 & $\pm$0.40 & $\pm$0.17 &            &           &           &           &       & \\ [6.0pt]

TT Ari    &      63.57 &     44.37 &     29.70 &      1.67 &     15.77 &     11.46 &      1.18 &      7.63 &      5.10 &   $<$0.40 &  \nodata   &   $<$1.83 &      0.90 & \nodata   & I~I~- & 3 \\ 
          &  $\pm$1.71 & $\pm$1.28 & $\pm$0.81 & $\pm$0.04 & $\pm$0.71 & $\pm$0.52 & $\pm$0.03 & $\pm$0.34 & $\pm$0.29 &           &            &           & $\pm$0.13 &           &       & \\ 
          &            &           &           &   [18.2]  &           &           &    [9.85] &           &           & [$<$1.8]  &            & [$<$2.2]  &           &           &       & \\ [6.0pt]

RW Tri    &      26.75 &     23.94 &     17.38 &      7.89 &      7.70 &      5.53 &      4.94 &      3.75 &      2.59 &      1.05 &  \nodata   &   $<$2.61 &      0.45 & \nodata   & I~I~- & 3 \\
          &  $\pm$0.74 & $\pm$0.71 & $\pm$0.48 & $\pm$0.21 & $\pm$0.35 & $\pm$0.25 & $\pm$0.12 & $\pm$0.17 & $\pm$0.12 & $\pm$0.15 &            &           & $\pm$0.09 &           &       & \\
          &            &           &           &           &           &           &           &           &           &           &            &           &    [0.5]  &           &       & \\ [6.0pt]

V347 Pup  &       8.93 &     10.42 &      8.10 &      6.92 &      7.24 &      5.09 &      4.08 &      3.65 &      2.36 &      0.97 &       0.76 &   $<$1.73 & \nodata   & \nodata   & I~I~- & 3 \\
          &  $\pm$0.27 & $\pm$0.31 & $\pm$0.27 & $\pm$0.18 & $\pm$0.33 & $\pm$0.23 & $\pm$0.10 & $\pm$0.17 & $\pm$0.14 & $\pm$0.07 &  $\pm$0.04 &           &           &           &       & \\
          &            &           &           &    [8.1]  &           &           &    [4.7]  &           &           &    [1.1]  &            & [$<$1.8]  &           &           &       & \\ [6.0pt]

UX UMa    &      12.57 &     11.13 &      8.29 &      5.99 &      5.42 &      3.88 &      3.76 &      2.73 &      1.69 &      0.82 &       0.51 &   $<$2.22 & \nodata   &      0.30 & I~I~I & 3 \\ 
          &  $\pm$0.37 & $\pm$0.39 & $\pm$0.25 & $\pm$0.16 & $\pm$0.25 & $\pm$0.18 & $\pm$0.09 & $\pm$0.12 & $\pm$0.08 & $\pm$0.10 &  $\pm$0.02 &           &           & $\pm$0.04 &       & \\ 
          &    [18.1]  &   [14.2]  &   [10.1]  &           &           &           &           &           &           &           &            &           &           &           &       & \\ [6.0pt]

VY Scl    &      11.66 &      7.84 &      5.55 &      2.69 &      1.03 &      0.78 &      1.67 &      0.56 &      0.36 &   $<$0.57 &      0.050 &   $<$2.02 & \nodata   & \nodata   & I~I~- & 3 \\
          &  $\pm$0.31 & $\pm$0.26 & $\pm$0.17 & $\pm$0.07 & $\pm$0.05 & $\pm$0.04 & $\pm$0.05 & $\pm$0.03 & $\pm$0.02 &           & $\pm$0.004 &           &           &           &       & \\
          &            &           &           &           &    [2.4]  &    [1.6]  &           &    [1.1]  &    [0.6]  &           &     [0.15] &           &           &           &       & \\ [6.0pt]

V3885 Sgr &     166.14 &    130.71 &     94.97 &     50.90 &     48.29 &     34.39 &     30.63 &     24.43 &     15.35 &      7.45 & \nodata    &   $<$2.95 & \nodata   & \nodata   & I~-~- & \nodata \\
          &  $\pm$4.96 & $\pm$4.46 & $\pm$2.84 & $\pm$1.41 & $\pm$2.17 & $\pm$1.59 & $\pm$0.78 & $\pm$1.11 & $\pm$0.70 & $\pm$0.24 &            &           &           &           &       & \\ [6.0pt]
\tablebreak
V442 Oph  &       7.43 &      5.23 &      3.78 &      0.88 &      1.15 &      0.94 &      0.54 &      0.62 &      0.56 &   $<$0.30 & \nodata    &   $<$2.15 & \nodata   & \nodata   & I~-~- & 3 \\
          &  $\pm$0.22 & $\pm$0.20 & $\pm$0.16 & $\pm$0.03 & $\pm$0.05 & $\pm$0.04 & $\pm$0.03 & $\pm$0.03 & $\pm$0.05 &           &            &           &           &           &       & \\
          &            &           &           &    [1.3]  &           &           &    [0.75] &           &           & [$<$0.3]  &            & [$<$2.2]  &           &           &       & \\ [6.0pt]

WX Ari    &       2.74 &      2.35 &      1.80 &      1.02 &      1.07 &      0.78 &      0.70 &      0.57 &      0.44 &   $<$0.58 & \nodata    &   $<$2.89 & \nodata   & \nodata   & \nodata & \nodata \\
          &  $\pm$0.10 & $\pm$0.12 & $\pm$0.09 & $\pm$0.03 & $\pm$0.05 & $\pm$0.04 & $\pm$0.03 & $\pm$0.03 & $\pm$0.03 &           &            &           &           &           &         & \\ [6.0pt]

\enddata
\tablenotetext{a}{All flux densities are given in mJy and wavelengths in $\mu$m.}
\tablenotetext{b}{Methods used to extract the tabulated Spitzer photometry for IRAC, IRSPUI, and MIPS: I = IRAF aperture photometry, A = MOPEX/APEX aperture photometry, P = MOPEX/APEX PRF-fitting photometry.}
\tablecomments{
               (1) 2MASS photometry given in brackets is dereddened as described in \citet{hoard09}, while the reddending corrections at longer wavelengths are smaller than the photometric uncertainties; 
               (2) W3 flux density given in brackets is the PRF-fit value derived in this work; 
               (3) flux density values given in brackets are offset-corrected values discussed in this work.
              }
\end{deluxetable}

%% file: table_corrs.tex
\begin{deluxetable}{lllllll}
\tablewidth{0pt}
\tabletypesize{\scriptsize}
\tablecaption{System Parameter Correlations \label{t:corrs}} 
\tablehead{
\colhead{Case\tablenotemark{a}} & 
\colhead{Parameter} & 
\colhead{$\lambda_{\rm excess}$} & 
\multicolumn{2}{l}{Linear Fit:} &
\multicolumn{2}{l}{Spearman's Rank:} 
\\
\colhead{ } & 
\colhead{ } & 
\colhead{($\mu$m)} & 
\colhead{slope} &
\colhead{$\tilde\chi^{2}$} &
\colhead{$\rho$} & 
\colhead{$P$\tablenotemark{b}}
}
\startdata
 A & Inclination    & 5.7 & \phs$0.0027(8)$ deg$^{-1}$  & 0.8 & \phs$0.65$ & 0.02 \\
   &                & 7.9 & \phs$0.0056(13)$ deg$^{-1}$ & 3.9 & \phs$0.62$ & 0.03 \\ [2pt]
   & Orbital Period & 5.7 & $-0.22(22)$ d$^{-1}$        & 2.1 & $-0.29$    & 0.35 \\
   &                & 7.9 & $-0.26(27)$ d$^{-1}$        & 6.6 & $-0.36$    & 0.25 \\ [6pt]

 B & Inclination    & 5.7 & \phs$0.0022(9)$ deg$^{-1}$  & 0.5 & \phs$0.65$ & 0.04 \\
   &                & 7.9 & \phs$0.0044(13)$ deg$^{-1}$ & 1.7 & \phs$0.55$ & 0.10 \\ [2pt]
   & Orbital Period & 5.7 & \phs$0.03(23)$ d$^{-1}$     & 1.6 & $-0.03$    & 0.93 \\
   &                & 7.9 & \phs$0.07(27)$ d$^{-1}$     & 3.9 & $-0.12$    & 0.75 
\enddata
\tablenotetext{a}{Case A includes all of the targets, while case B excludes the outliers WX~Ari and V442~Oph.}
\tablenotetext{b}{$P$ is the significance of the correlation (i.e., the probability that it could have occurred by chance).}
\end{deluxetable}

%% file: table_models_part1.tex
\begin{deluxetable}{lllll}
\tablewidth{0pt}
\tabletypesize{\scriptsize}
\rotate
\tablecaption{Model Parameters for Selected Targets \label{t:models1}} 
\tablehead{
}
\startdata
Star                                    & V592 Cas                  & IX Vel                     & UX UMa                                                & RW Sex \\ 
T$_{\rm wd}$ (kK)                       & 45 [1,2]                  & 60(10) [1,3]               & 20 [1,4]                                              & 50 [1,5] \\ 
M$_{\rm wd}$ (M$_{\odot}$)              & 0.75 [1,2]                & 0.8(2) [1,3]               & 0.47 [1,4,6], $(0.83$--$1.01)\pm0.20$ [7]             & 0.90 [1,5] \\ 
R$_{\rm wd}$ (R$_{\odot}$)              & 0.0106 [1,2]              & 0.0150 [1,3]               & 0.0165 [1,4], 0.0139 [6]                              & 0.00882 [1,5] \\ 
Sp[2]                                   & M4.1 [1,8]                & M2.6 [1,8]                 & M2.6 [1,8]                                            & $\approx$M1 [1,8] \\ 
T$_{2}$ (K)                             & 3293 [1,8]                & 3500(1000) [3], 3614 [1,8] & 3623 [1,8]                                            & $\gtrsim$3908 [1,8] \\ 
M$_{2}$ (M$_{\odot}$)                   & 0.200 [1,8]               & 0.52(10) [3], 0.448 [1,8]  & 0.47 [4,6], $(0.36$--$0.43)\pm0.10$ [7], 0.459 [1,8]  & $\gtrsim$0.599 [1,8], 0.674 [5] \\ 
\.{M} ($10^{-9}$ M$_{\odot}$ yr$^{-1}$) & 15 [1,2]                  & $5.0\pm1.0$ [3], 7.1 [1,8] & 5--10 [4], 6--16 [6], 8.0 [1]                         & 2.0 [1,5], 5.75 [5] \\ 
R$_{\rm acd, in}$ (R$_{\rm wd}$)        & 1.00 [1,2]                & 1.00 [1,3]                 & 1.00 [1,4]                                            & 1.00 [1,5] \\ 
R$_{\rm acd, out}$ (R$_{\rm wd}$)       & 35.00 [1,2]               & 37.33 [1,3]                & 29.58 [1,4], 29.90 [6]                                & 40.00 [1], 50.00 [5] \\ 
h$_{\rm acd, in}$ (R$_{\rm wd}$)        & 0.05 [1,2]                & 0.178 [1]                  & 0.039 [1]                                             & 0.05 [1] \\ 
h$_{\rm acd, out}$ (R$_{\rm wd}$)       & 1.68 [1,2]                & 13.33 [1,3]                & 0.96 [1,4]                                            & 4.2 [1] \\ 
f$_{\rm acd,3.55}$ (\%)                 &   93                      & 78                         & 54                                                    & 50   \\ 
f$_{\rm acd,7.87}$ (\%)                 &   77                      & 69                         & 42                                                    & 38   \\ 
f$_{\rm acd,23.68}$ (\%)                &   26                      & 53                         & 26                                                    & 29   \\ 

T$_{\rm cbd, in}$ (K)                   & 500 [1,2]                 & 1000 [1]                   & 1000 [1]                                              & 800 [1] \\ 
T$_{\rm cbd, out}$ (K)                  &  50 [1,2]                 &   20 [1]                   &   20 [1]                                              & 20 [1] \\ 
R$_{\rm cbd, in}$ (R$_{\rm wd}$)        & 700 [1,2]                 &  172 [1]                   &   143 [1]                                             & 352 [1] \\ 
R$_{\rm cbd, out}$ (R$_{\rm wd}$)       & 15000 [1,2]               & 31677 [1]                  & 26344 [1]                                             & 48110 [1] \\ 
M$_{\rm cbd, total}$ ($10^{21}$~g)\tablenotemark{a} & 2.30 [1,2]    & 1.29 [1]                   & 1.40 [1]                                              & 1.1 [1] \\ 
f$_{\rm cbd,3.55}$ (\%)                 & $<$1                      &  1                         &  3                                                    & $<$1 \\ 
f$_{\rm cbd,7.87}$ (\%)                 &   15                      & 10                         & 18                                                    &  7   \\ 
f$_{\rm cbd,23.68}$ (\%)                &   72                      & 33                         & 51                                                    & 32   \\ 
\enddata
\tablenotetext{a}{Total dust mass corresponds to grains with radius of $r_{\rm grain}=1$~$\mu$m and scales with $r_{\rm grain}$.}
\tablecomments{The f$_{{\rm acd,}\lambda}$ and f$_{{\rm cbd,}\lambda}$ parameters are the fraction of total modelled system light contributed by the accretion disk and circumbinary disk components, respectively, at the indicated wavelength (corresponding to IRAC-1, IRAC-4, or MIPS-24).}
\tablerefs{~Listed in brackets with each parameter value: 
1 = this work; 
2 = \citet{hoard09};
3 = \citet{linnell07};
4 = \citet{linnell08a};
5 = \citet{linnell10} -- the smaller (larger) \.{M} corresponds to the smaller (larger) distance estimate listed in Table~\ref{t:targets};
6 = \citet{noebauer10};
7 = \citet{neustroev11} -- the larger component mass estimates correspond to the lower inclination from Table~\ref{t:targets};
8 = \citet{knigge11}.
}
\end{deluxetable}

%% file: table_models_part2.tex
\begin{deluxetable}{lllllllll}
\tablewidth{0pt}
\tabletypesize{\scriptsize}
\setlength{\tabcolsep}{0.03in} 
\rotate
\tablecaption{Published System Parameters \label{t:models2}} 
\tablehead{
}
\startdata
Star                                    & QU Car                      & TT Ari                   & RW Tri                                       & V347 Pup                        \\ 
T$_{\rm wd}$ (kK)                       & 55 [1]                      & \nodata                  & \nodata                                      & \nodata                         \\ 
M$_{\rm wd}$ (M$_{\odot}$)              & 0.60, 1.20 [1]              & 1.24(20) [2]             & 0.4--0.7 [3], 0.7(1) [4,5], 0.55 [6]         & 0.63(4) [7]                     \\ 
R$_{\rm wd}$ (R$_{\odot}$)              & 0.0159, 0.00666 [1]         & \nodata                  & 0.0115 [5]                                   & 0.0112(1) [8]                   \\ 
Sp[2]                                   & $<$M1 [9]                   & M4.1 [9]                 & M1.5 [9]                                     & M0.5V [7], K0--5V [8], M1.5[9] \\ 
T$_{2}$ (K)                             & 3500, 6500 [1], $>$3900 [9] & 3311 [9]                 & 3892 [9]                                     & 3894 [9]                       \\ 
M$_{2}$ (M$_{\odot}$)                   & 0.50, 1.2 [1], $>$0.6 [9]   & 0.23(2) [2], 0.208 [9]   & 0.3--0.4 [3], 0.6 [4,5], 0.35 [6], 0.594 [9] & 0.52(6) [7], 0.594 [9]         \\ 
\.{M} ($10^{-9}$ M$_{\odot}$ yr$^{-1}$) & 100--1000 [1]               & 1 [10]                   & 10 [4], 2.7--26, 8 (preferred) [5], 4.6 [6]  & $\sim6$ [11], $\approx10$ [12]    \\ 
R$_{\rm acd, in}$ (R$_{\rm wd}$)        & 1.00, 1.00 [1]              & \nodata                  & 1.00 [5]                                     & \nodata                         \\ 
R$_{\rm acd, out}$ (R$_{\rm wd}$)       & 52.83, 149.85 [1]           & \nodata                  & 30.00 [5]                                    & 0.72(9) [7]                     \\ 
h$_{\rm acd, in}$ (R$_{\rm wd}$)        & \nodata                     & \nodata                  & \nodata                                      & \nodata                         \\ 
h$_{\rm acd, out}$ (R$_{\rm wd}$)       & 3.14, 7.51 [1]              & \nodata                  & \nodata                                      & \nodata                         \\ [10pt] 

Star                                    & V3885 Sgr                   & VY Scl                   & V442 Oph                                     & WX Ari \\ 
T$_{\rm wd}$ (kK)                       & 57(5) [13]                  & 45(3) [14]               & 23 [12]                                      & \nodata \\ 
M$_{\rm wd}$ (M$_{\odot}$)              & 0.7(1) [13]                 & 1.0 [14], 1.22(22) [15]  & 0.4 [12]                                     & $\approx0.8$ [16], 0.35 [12] \\ 
R$_{\rm wd}$ (R$_{\odot}$)              & 0.0134 [13]                 & \nodata                  & \nodata                                      & \nodata \\ 
Sp[2]                                   & M2.2 [9]                    & M1.5 [9]                 & M4.0 [9]                                     & M4.1 [9] \\ 
T$_{2}$ (K)                             & 3650(100) [13], 3698 [9]    & 3896 [9]                 & 3289 [9]                                     & 3319 [9] \\ 
M$_{2}$ (M$_{\odot}$)                   & 0.475 [13], 0.504 [9]       & 0.43(13) [15], 0.595 [9] & 0.200 [9]                                    & $\approx0.31$ [16], 0.306 [9] \\ 
\.{M} ($10^{-9}$ M$_{\odot}$ yr$^{-1}$) & 5(2) [13]                   & 8 [14]                   & 10 [12]                                      & 1 [12,16] \\ 
R$_{\rm acd, in}$ (R$_{\rm wd}$)        & 1.00 [13]                   & 1.05 [14]                & \nodata                                      & \nodata \\ 
R$_{\rm acd, out}$ (R$_{\rm wd}$)       & 41.79 [13]                  & \nodata                  & \nodata                                      & \nodata \\ 
h$_{\rm acd, in}$ (R$_{\rm wd}$)        & \nodata                     & \nodata                  & \nodata                                      & \nodata \\ 
h$_{\rm acd, out}$ (R$_{\rm wd}$)       & 7.46 [13]                   & \nodata                  & \nodata                                      & \nodata \\ 
\enddata
\tablerefs{~Listed in brackets with each parameter value:
1 = \citet{linnell08b} -- the higher mass WD model is preferred, but actual system parameters could lie inside the range of parameter values;
2 = \citet{wu02};
3 = \citet{poole03};
4 = \citet{groot04};
5 = \citet{noebauer10};
6 = \citet{puebla11};
7 = \citet{thoroughgood05};
8 = \citet{diaz99};
9 = \citet{knigge11};
10 = \citet{ballouz09};
11 = \citet{puebla07};
12 = \citet{ballouz09};
13 = \citet{linnell09};
14 = \citet{hamilton08};
15 = \citet{pais00};
16 = \citet{gil00}.
}
\end{deluxetable}